\documentclass[twocolumn]{aastex701}

\newcommand{\msun}{\ensuremath{M_\sun}}

\newcommand{\chimera}{{\sc Chimera}}
\newcommand{\isotope}[2]{\ensuremath{\mathrm {^{#2}#1}}}

\newcommand{\kmps}{\ensuremath{\mbox{km~s}^{-1}}}

\newcommand{\gcc}{\ensuremath{{\mbox{g~cm}}^{-3}}}
\newcommand{\colden}{\ensuremath{{\mbox{cm}}^{-2}}}
\newcommand{\nitr}{\isotope{Ni}{56}+Tr}
\newcommand{\vlos}{\ensuremath{v_{\rm LOS}}}
\newcommand{\mNS}{\ensuremath{M_{\rm NS}}}

\newcommand{\runlabel}[3]{\textnormal{#1}\textit{#2}\textnormal{#3}}

\defcitealias{Sandoval_2021}{MAS+21}

\begin{document}

\title{Simulation to a Newborn Supernova Remnant from a Low-mass Iron Core Star}

\author[orcid=0000-0003-3944-3645,sname='Neopane']{Sudarshan Neopane}
\affiliation{Department of Physics and Astronomy, University of Tennessee, Knoxville, TN 37996-1200, USA}
\email[show]{sneopane@vols.utk.edu}  

\author[orcid=0000-0002-5088-4487, sname='Sandoval']{Michael A. Sandoval} 
\affiliation{National Center for Computational Sciences, Oak Ridge National Laboratory, P.O. Box 2008, Oak Ridge, TN 37831-6164, USA}

\email{sandovalma@ornl.gov}

\author[orcid=0000-0002-9481-9126, gname=Raph,sname=Hix]{W. Raphael Hix}
\affiliation{Physics Division, Oak Ridge National Laboratory, P.O. Box 2008, Oak Ridge, TN 37831-6354, USA}
\affiliation{Department of Physics and Astronomy, University of Tennessee, Knoxville, TN 37996-1200, USA}
\email{raph@utk.edu}

\author[orcid=0000-0003-3023-7140]{J. Austin Harris}
\affiliation{National Center for Computational Sciences, Oak Ridge National Laboratory, P.O. Box 2008, Oak Ridge, TN 37831-6164, USA}
\email{harrisja@ornl.gov}

\author[orcid=0000-0002-5358-5415]{O. E. Bronson Messer}
\affiliation{National Center for Computational Sciences, Oak Ridge National Laboratory, P.O. Box 2008, Oak Ridge, TN 37831-6164, USA}
\affiliation{Physics Division, Oak Ridge National Laboratory, P.O. Box 2008, Oak Ridge, TN 37831-6354, USA}
\affiliation{Department of Physics and Astronomy, University of Tennessee, Knoxville, TN 37996-1200, USA}
\email{bronson@ornl.gov}

\author[/0000-0002-5231-0532]{Eric J. Lentz}
\affiliation{Department of Physics and Astronomy, University of Tennessee, Knoxville, TN 37996-1200, USA}
\affiliation{Physics Division, Oak Ridge National Laboratory, P.O. Box 2008, Oak Ridge, TN 37831-6354, USA}
\email{elentz@utk.edu}

\correspondingauthor{Sudarshan Neopane}

\begin{abstract}

Supernova remnant observations show a high degree of asymmetry, mixing, and inhomogeneity. 
These asymmetries are seeded during the early seconds of the explosion and are further enhanced and modified as the shock and ejecta move through the stellar progenitor and into the circumstellar medium. 
We present simulations of a 9.6~\msun\ zero-metallicity progenitor initialized after shock revival and evolved for several years when the ejecta is in the circumstellar medium. 
A suite of 1D and 2D simulations examines the effects of neutron-star wind and radioactive decay heating. 
In 1D, decay heating forms a low-density bubble that suppresses the reverse shock. 
While in 2D, the heating is localized to metal-rich pockets, inflating them and compressing the surrounding material into dense shells. 
In 3D the neutron-star wind and decay heating modify the plume morphology, producing more large-scale structures. 
The extended plume morphology leads to an asymmetrical shock breakout. 
After breakout, the leading plumes cannot keep up with the shock front, resulting in deceleration and fragmentation by the reverse shock while retaining the large-scale asymmetry. 
The projected ejecta morphology and velocities are strongly viewing angle dependent.
The relatively uniform metal-rich distribution does not resemble the strongly inhomogeneous ejecta structure of Cas~A. The 160-isotope decay network shows that 24.4\% of the radioactive heating comes from decay chains other than the canonical \isotope{Ni}{56} chain.
The low explosion energy, low \isotope{Ni}{56} yield, and Ni/Fe ratio greater than unity suggest an observational signature similar to an electron capture supernova.

\end{abstract}



\section{Introduction} 
\label{sec:intro}

Core-collapse supernovae (CCSNe) are explosions of massive stars whose cores can no longer support their own gravitational mass.
These explosions enrich the interstellar medium with elements ranging from oxygen to iron and some elements heavier than iron, and thus play a key role in galactic chemical evolution \citep{Tsujimoto2022, Trimble1991}. A broad set of observations shows that these explosions are highly asymmetric and strongly mixed, with newly synthesized material emerging in clumps, knots, and high-velocity ``bullets" rather than smooth, spherical shells.
The high ejection velocities of iron-group elements and $^{56}$Ni ``bullets" inferred for SN~1987A point to strong outward mixing of radioactive material \citep{Boggs_2015}. 
Early photospheric emission \citep{Utrobin_2015, Utrobin_2017} and nebular-phase spectral analyses \citep{Jerkstrand2017} likewise require substantial large-scale mixing to reproduce line profiles and luminosity evolution. 
Similarly, detections of freshly synthesized radioactive \isotope{Ti}{44} in the Cassiopeia A (Cas A) supernova remnant reveal an inhomogeneous, 3D distribution of the innermost ejecta \citep{Grefenstette_2017, Grefenstette_2014}.
These observed asymmetries are thought to originate from asymmetries in the central engine and are further amplified by hydrodynamic instabilities as the shock and ejecta propagate through the stellar interior, break out of the stellar surface, and expand into the circumstellar medium (CSM). 
The need to explain clumpy, bullet-like structures and strong mixing has driven decades of multidimensional studies of shock propagation in CCSNe. 
In this paper, we present a state-of-the-art 3D CCSN simulation that follows the growth and evolution of asymmetries as the shock traverses the progenitor star and expands into the CSM over years post-bounce.

Early hints of asymmetry, such as fine structure in early-time line profiles, spectropolarimetric signatures, and evidence for high-velocity \isotope{Ni}{56} ``bullets'' in SN~1987A, motivated the first attempts (in 2D) to follow shock propagation beyond strict spherical symmetry \citep{Hachisu1990, Mueller1991, HerantBenz1992}. 
An early attempt at 3D was models performed in spherical symmetry until the blast wave had reached the (C+O)/He or He/H interface, then mapped to multidimensional grids with added perturbations to the radial velocity field \citep{Kane_2000}. 
These perturbations broke the spherical symmetry and triggered the growth of instabilities intended to reproduce the observed clumps and bullets. 
However, there remained a significant discrepancy in the \isotope{Ni}{56} mixing between the observed and simulated ejecta of SN~1987A, particularly in the extent and velocities of nickel-rich structures. Despite being in axisymmetry, the neutrino-driven simulations of \citet{kifiondis2003} showed significant differences in the ejecta compared to previous parameterized models, illustrating the importance of realistic stellar density profiles and full multidimensional evolution of the shock. Neutrino-heating–powered series of 2D and 3D shock breakout models by \citet{Hammer2010} further demonstrated that Rayleigh–Taylor (RT) instabilities, and hence the development of \isotope{Ni}{56} clumps and bullets, proceed much more slowly in 2D than in 3D, in agreement with \citet{Kane_2000}.

Improving upon the work by \cite{Hammer2010},  \cite{Wongwathanarat2015} simulated four different progenitors, three 15~\msun\ and one 20~\msun\, in 3D with full coverage of the progenitor. They used a subset of \cite{Anop2013} neutrino-driven explosion as their initial condition for these simulations. Their simulations implemented two analytical spherically symmetric wind prescriptions: constant and a time-dependent power law wind. In addition to the wind, they used ideal gas law with radiation contributions for regions where the density and temperature value fall below the tabulated values of the equation of state. For regions outside the progenitor star, a steady stellar wind approximation was used. The simulations were run until shock breakout revealing that, even though the clumpiness of the iron group elements is related to the initial asymmetries, the morphology of the ejecta, which is determined by the shock and reverse shock dynamics, is tied to the progenitor density profile. \cite{gabler_2020} continued some of these simulations to times longer than 1 year after core bounce.
They found that the decay heating accelerates the iron-rich ejecta and found volume filling factors for three of four models consistent with the observations of SN~1987A. In the followup work to \cite{Wongwathanarat2015} by \cite{Wongwathanarat2017}, a red supergiant (RSG) model was modified by removing most of the hydrogen envelope to understand relations between explosion asymmetries and observable ejecta features in Cas A. Both \cite{Wongwathanarat2015} and \cite{Wongwathanarat2017} simulations were performed using the PROMETHEUS-HOTB code, which has a three-flavour gray neutrino transport scheme and a 15 isotope network with an additional tracer species. 

Similarly, \cite{stockinger2020}
have run a series of 1D, 2D, and 3D shock breakout simulations of low mass progenitors consisting of one ONeMg core with mass 8.8~\msun\ and two iron cores with masses  9~\msun\ and 9.6~\msun. Their extensive study includes full 3D runs extending from core collapse through shock breakout and multi-day fallback. Their results demonstrate a strong causal link between the pre-collapse progenitor structure and the subsequent explosion morphology, mixing behavior, and nucleosynthetic outcomes. In particular, the steep density gradients of low-mass (8--10 \msun) progenitors promote early shock revival and drive highly asymmetric explosions, producing large-scale RT plumes that penetrate through multiple composition interfaces. These instabilities mix Fe-group nuclei and $\alpha$-rich freeze-out products into the outer layers, while also channeling anisotropic fallback that influences the neutron star's spin and kick. 
While their ONeMg core model utilized PROMETHEUS-HOTB for collapse, the iron core models used VERTEX-PROMETHEUS by \cite{Melson2015a, Melson2020}, which includes a three-flavor, three-species energy-dependent, ray-by-ray-plus (RbR+) neutrino transport scheme with a 15-isotope nuclear reaction network.

A series of five 3D spherical CCSN simulations of low-mass progenitors, with masses 9.6~\msun\ and 10~\msun, until shock breakout were also been performed by \cite{Sandoval_2021} (hereafter \citetalias{Sandoval_2021}). 
Their study improved on the previous study by including a more realistic 160-isotope nuclear reaction network. 
They used two 2D and one 3D explosion models and mapped them to 3D for evolution. 
To test the utility of a 2D model in the absence of a complete 3D model, \citetalias{Sandoval_2021} tilted the two 2D progenitor models by 90$^\circ$ and evolved them in 3D to remove the alignment of 2D and 3D axes. 
They found that tilting the $y$-axis of a 2D progenitor is a viable and much better method to study instabilities and mixing in the absence of a complete 3D model. However, it doesn't produce all the features present in a 3D model and therefore, it is not a perfect replacement. 
They also compared the ejecta yields of the same low-mass progenitor with those from \cite{stockinger2020}, which had lower explosion energy and used a 15-isotope network with tracer species. 
Comparison by \citetalias{Sandoval_2021} showed that the power of explosion energy, and hence accurate modeling of the central engine,  can radically affect ejecta morphology. 
Further, \citetalias{Sandoval_2021} showed that the tracer species don't fully capture the yields of neutron-rich isotopes at shock breakout, and the size of the nuclear network used can be crucial in determining the ejecta distribution. The collapse models used by \citetalias{Sandoval_2021} were generated using the \chimera\ code, which uses four-species, three-flavor,  energy-dependent, RbR+ neutrino transport scheme with a 160-isotope nuclear reaction network.

More recently, \cite{Vartanyan_2025} have run a 17-\msun\ CCSN model through a pipeline that connects the core-collapse phase to the observable post-explosion evolution, extending from the neutrino-driven engine to the hydrodynamic shock breakout and radiative emission. Their end-to-end approach showed that early asymmetries and RT mixing persist through breakout, shaping the broadband and bolometric light curves. Additionally, in \cite{vartanyan2025_largesims}, a series of six progenitors models with masses ranging from 9~\msun\ to 25~\msun\ were ran until shock breakout and through the homologous expansion of the CSM. They find strongly asymmetric, clumpy nickel ejecta whose large-scale dipole pattern is established in the first seconds after explosion and largely preserved out to breakout, with substantial mixing of \isotope{Ni}{56} into the hydrogen envelope and a tight correlation between early dipole asymmetry, mixing, and nickel velocities. The collapse simulations were performed using the FORNAX code \citep{Skinner_2019}, which includes a three-flavor energy-dependent neutrino transport scheme. The nucleosynthesis from the simulations were post-processed rather than computed \textit{in situ}.

Finally, \cite{giudici2025_shock_breakout} carried out long-term 3D neutrino-driven simulations of Type IIP supernovae for 13 RSG progenitors (12.5--27.3~\msun), evolved from core bounce, using the PROMETHEUS-HOTB code, to 10 days post-explosion, to study how progenitor structure controls macroscopic mixing. They show that RT instabilities that grow at the (C+O)/He and He/H interfaces, and their interaction with the reverse shock at He/H, are the key drivers of outward \isotope{Ni}{56} transport into the hydrogen envelope. The most efficient Ni mixing and highest Ni velocities actually occur in lower-mass RSGs, while the most massive models exhibit weak mixing, and the radial Ni-mixing efficiency anti-correlates linearly with He-core mass and correlates with a local maximum of $\rho r^{3}$ in the He shell, where shock deceleration boosts RT growth. These correlations provide a physically motivated way to prescribe macroscopic mixing in 1D light-curve models and suggest that observed degrees of Ni mixing in Type IIP SNe can be used to infer internal RSG structure.

Building on these advances and motivated by the growing observational evidence that the most diagnostic asymmetries are set early but reshaped by interface instabilities and shock interaction with the outer envelope and CSM, we use a single, fully 3D state-of-the-art simulation to connect the neutrino driven explosion phase to the multi year hydrodynamic evolution of the ejecta. The paper is organized as follows. 
Section 2 outlines our methodology, including the grid setup and numerical scheme, the progenitor and collapse model used to initialize the explosion, the parametrized PNS mass-loss and wind prescription, the circumstellar medium structure, and our treatment of nuclear burning and radioactive decay through the reaction network. 
Then, in Section~3, we present the results of our 1D and 2D simulations, with particular emphasis on the influence of the additional physics. 
In Section~4, we describe the 3D simulation and potential observable signatures. 
Finally, we summarize in Section~5 our main conclusions and discuss the implications for interpreting observed asymmetries in supernovae and their remnants.

\section{Methodology} \label{sec:methodology}
Our simulations are performed using the Flash-X code \citep{dubey2022}, a successor of the widely used FLASH code \citep{dubey2009, Fryxell_2000}.
Flash-X is a multiphysics code developed to solve problems formulated in ordinary and partial differential equations on heterogeneous architectures. The gravity and nuclear kinetic coupled Euler equations are solved using the Spark \citep{dubey2022} hydrodynamic solver. Spark uses the WENOz5 \citep{couch2021, dubey2022} reconstruction scheme for fifth-order reconstruction, the HLL \citep{Toro2009} and HLLC solver to compute the Riemann fluxes across the cell interfaces, and the 2nd order Strong Stability Preserving (SSP) Runge-Kutta to integrate in time. We find that the WENOz5 reconstruction is prone to Gibbs phenomenon near strong compositional and density gradients.
To work around this issue, we revert to TVD reconstruction \citep{Toro2009} in regions that do not preserve monotonicity between the cell-centered values and the face reconstructed values.

\subsection{Grid Setup} \label{sec:grid_setup}
We utilized spherical geometry with the modified uniform grid (UG) for our simulations. Unlike the regular UG, which keeps the cell width the same in all directions, our modified grid keeps $\Delta r/r$ roughly constant in the radial direction. Other spatial directions are unchanged. This modification allows us to incorporate large radial extents in our domain while keeping the block count, where each block is a collection of compute units called cells, computationally feasible \citep{Fernandez2012}. This grid setup has been used for numerous shock breakout simulations \citep{Sandoval_2021, stockinger2020, Vartanyan_2025}. To achieve simulation times of the order of years, it is necessary to remove the neutron star (NS), with characteristic timescale $< 1$~s, from the domain. 
Since we do not include neutrino transport in our simulation, we are unable to accurately follow the evolution of the NS and its immediate environment. We therefore cut the inner 500~km and approximate the mass in the excised inner 500~km region as a point mass for gravitational potential computation. This point mass is also considered as the mass of the 
NS. As the simulation progresses, the sound-speed at the inner regions of the domain limits the time step, therefore to make the simulations run faster, we progressively drop blocks from the inner region as the shock moves outward.
We have also removed a wedge of $5^\circ$ around the poles in the $\theta$ direction, similar to \citet{Sandoval_2021} and  \citet{vartanyan2025_largesims, Vartanyan_2025}  to further alleviate the CFL time constraint in the $\phi$ direction.
The mass in the wedge region is also considered as a point mass at the origin for gravitational potential computation. We have the ability to extend the domain by adding blocks in the radial direction, which allows tracking of the evolution of shock and ejecta for much longer timescale. This ability to extend the grid is crucial in making the simulations into the circumstellar medium (CSM) computationally feasible. 

\subsection{NS wind, mass loss and accretion} \label{sec:pns_wind_cool}
As mentioned above, our simulation does not have a NS at the center, instead, we consider the region in the cutout to be a point-mass representing the mass of the NS. 
Our simulation includes a spherically symmetric NS power-law wind, same as \cite{Wongwathanarat2015}. 
The wind is applied in the inner radial boundary with the wind density ($\rho_{w}$), internal energy ($e_{w}$), and temperature ($T_{w}$) following a power law given by
\begin{equation}
    \rho_{w}(t) = \rho_{w}(t_{\rm map}) \times \left(\frac{t}{t_{\rm map}} \right)^{-7/2},
\end{equation}
\begin{equation}
e_{w}(t) = e_{w}(t_{\rm map}) \times \left(\frac{t}{t_{\rm map}} \right)^{-7/6},    
\end{equation}
\begin{equation}
    T_{w}(t) = T_{w}(t_{\rm map}) \times \left(\frac{t}{t_{\rm map}} \right)^{-7/6}
\end{equation}
where, $t_{map}$ is the time of mapping to the Flash-X grid and the wind pressure is computed using these new density and internal energy values as
\begin{equation}
\label{eqn:press}
    p_{w}(t) = \frac{1}{3} e_{w}(t) \rho_{w}(t)
\end{equation}
The wind properties at the time of mapping, except for the pressure, are obtained by averaging their respective values in the row of cells adjacent to the inner boundary. 

Numerically, this prescription is imposed by filling the ghost cells at the inner radial boundary with the prescribed wind state. 
The mass, momentum, and energy fluxes entering the computational domain are then determined by the Riemann solve at the boundary interface.
To be consistent with \cite{Wongwathanarat2015}, we replace the equation-of-state call in the ghost cells with Equation~(\ref{eqn:press}). 

The NS wind prescription is expected to affect the innermost ejecta, the wind inflated cavity and the early compression of the material near the base of the ejecta.
Multidimensional neutrino radiation hydrodynamic simulations show that neutrino-driven winds can be deformed, aspherical, and channeled by the surrounding ejecta and fallback flows \citep[see, e.g.,][]{WangBurrows_2023}.
Ideally, extended CCSN models would be based on such simulations that complete the PNS wind phase.
However, this is not always possible, in which case it must be acknowledged that the detailed morphology of the cavity and its influence on the early development of RT plumes is likely simplified by the adopted spherical wind prescription.

We also consider mass loss from the NS due to neutrino emission using an  exponential neutrino cooling model similar to the one described in \cite{Fernandez_2018}. In our case, the mass loss $M_{\rm loss}$ due to neutrino emission of the NS of mass \mNS\ in time $\Delta t$ is:
\begin{equation}
    \mathrm{M}_{\rm loss}(t) = \frac{\mathrm{B.E.}(\mNS)}{\tau_{c}} \exp{\left(\frac{-(t-t_{\rm map})}{\tau_{c}}\right)} \Delta t  
\end{equation}
where, $\tau_{c}$ is a fiducial neutrino cooling time-scale, and \begin{equation}
    \mathrm{B.E.}(\mNS) = 0.084 \left(\frac{\mNS}{\msun}\right)^{2} \msun
\end{equation} 
is the gravitational binding energy of a cold neutron star \citep{ LatYahil1989, Lattimer_2001}.
Following \cite{Fernandez_2018}, we use $\tau_{c} = 3.0$~s.

During the simulation, as the matter flows out of the computational domain through the inner ``diode" boundary, it is considered to have accreted to the NS. Once we start dropping blocks, the in-fall mass from the new inner radial boundary doesn't necessarily get accreted directly to the NS, but is still included in the point mass at origin for the gravitational potential computation. 

\subsection{Circumstellar Medium (CSM)} \label{sec:csm_section}
To truly achieve our goal of running the simulations much beyond shock breakout, a physically realistic CSM profile is necessary. 
The CSM allows the ejecta time and space to expand, after it breaks out from the progenitor surface, i.e., becomes a supernova. 
The detailed structure of the circumstellar environment reflects both the environment in which the star formed and the changes made by the star's radiation and wind. 
In the absence of 3D models of the star's birth and circumstellar evolution, we adopt the approximation by  \citet{Wongwathanarat2015, Wongwathanarat2017}. 
The CSM is stationary and spherically symmetric, with density and temperature following a $r^{-2}$ decline from their value at the stellar radius, reflecting a steady stellar wind. 
The composition of the CSM is approximated to be the same as the edge of the progenitor star.
To initialize the CSM, density and temperature are used as inputs to the EOS.
The post-breakout evolution is naturally tied to this adopted circumstellar structure, with the rate at which the CSM is swept up, and hence the deceleration of the shock, directly dependent on the density profile of the CSM.
Additionally, an aspherical or time-dependent CSM could alter the angular dependence of shock propagation after breakout, the timing and strength of the reverse shock and the subsequent interaction of the reverse shock with the leading ejecta plumes.
Therefore, the detailed post-breakout plume deceleration and fragmentation should be interpreted in the context of the adopted CSM model.

\subsection{Equation of state (EOS)} \label{sec:eos}
We utilize the Flash-X implementation of the Helmholtz EOS \citep{Timmes_2000} to derive thermodynamic quantities from density and internal energy. The Helmholtz EOS includes contributions from radiation, completely ionized nuclei, and degenerate/relativistic electrons and positrons. The excision of the NS from the grid allows the density and temperatures in our simulation domain to stay under the upper limit of the EOS, which is only valid for $10^{-12} < \rho < 10^{15}$~\gcc\ and $10^{3} < T < 10^{13}$~K.
However, with the inclusion of a CSM, we run into issues with the lower limits of the EOS when the domain becomes sufficiently large. To avoid flooring of the thermodynamic values at the lower limits, we extended the lower limits of the EOS to $10^{-18}$~\gcc\ and $10^{-1}$~K. 
However, for the temperature, we assume a minimum value of 500~K in the CSM. With these new limits, we can now incorporate much larger domains, allowing tracking of the shock and ejecta for much larger timescales.

\subsection{Nuclear Reaction Network} \label{sec:nuclear_reaction}
Flash-X code comes with 2 fixed-sized nuclear reaction networks: approx-13 and approx-19 \citep{Timmes1999}, as well as a general-purpose nuclear reaction network XNet \citep{HixThelemann1999}. XNet is a stand-alone nuclear network code that can be generalized to larger reaction networks to improve the accuracy of species and energy yields. 
As in the \citetalias{Sandoval_2021} simulations, we utilize the 160-isotope \chimera/XNet reaction network  \texttt{sn160} to match the explosion phase \citep{LeHiHa26}, in order to track the detailed composition of the ejecta and adequately account for the feedback to hydrodynamics during our simulation.
\texttt{sn160} is an expanded version of the {\tt sn150} network \citep{Cher12,ChMeHi12,Harris_2017} extended to include Ga and Ge and better flow into and out of \isotope{Ca}{48}, the most neutron-rich isotope in the network.
The species included in \texttt{sn160} are \isotope{n}{}, \isotope{H}{1\textrm{--}2}, \isotope{He}{3\textrm{--}4}, \isotope{Li}{6\textrm{--}7}, \isotope{Be}{7,9},\isotope{B}{8,10,11}, \isotope{C}{12\textrm{--}14}, \isotope{N}{13\textrm{--}15}, \isotope{O}{14\textrm{--}18}, \isotope{F}{17\textrm{--}19}, \isotope{Ne}{18\textrm{--}22}, \isotope{Na}{21\textrm{--}23},\isotope{Mg}{23\textrm{--}26}, \isotope{Al}{25\textrm{--}27}, \isotope{Si}{28\textrm{--}32}, \isotope{P}{29\textrm{--}33}, \isotope{S}{32\textrm{--}36}, \isotope{Cl}{33\textrm{--}37}, \isotope{Ar}{36\textrm{--}40}, \isotope{K}{37\textrm{--}41}, \isotope{Ca}{40\textrm{--}48}, \isotope{Sc}{43\textrm{--}49}, \isotope{Ti}{44\textrm{--}51},  \isotope{V}{46\textrm{--}52}, \isotope{Cr}{48\textrm{--}54}, \isotope{Mn}{50\textrm{--}55}, \isotope{Fe}{52\textrm{--}58}, \isotope{Co}{53\textrm{--}59}, \isotope{Ni}{56\textrm{--}64}, \isotope{Cu}{57\textrm{--}65}, \isotope{Zn}{59\textrm{--}66}, \isotope{Ga}{62\textrm{--}64}, \isotope{Ge}{63\textrm{--}64}. 
Reaction rates are taken from the REACLIB\footnote{\url{https://groups.nscl.msu.edu/jina/reaclib/db/}} compilation (V2.0) \citep{CyAmFe10} and supplemented/supplanted with $\beta$-decay rates and electron capture rates on heavy nuclei \citep{FuFoNe85,OdHiMu94,LaMa00}.
 
In addition to the full nuclear reaction network, we also utilize a decay-only version of network that accounts for all the electron captures (ECs) and $\beta$-decays between the 160 isotopes with the isotopes rearranged in the solve to simplify and speed the numerical solution. 
For simplicity, we assume that 20\% of the decay energy is emitted in the form of neutrinos, which escape freely. 
This choice is motivated by the $\isotope{Ni}{56} \rightarrow \isotope{Co}{56} \rightarrow \isotope{Fe}{56}$ decay chain, for which the neutrino energy loss fraction is 18.98\% and 18.38\%, respectively \citep{Nadyozhin1994}.
The remaining decay energy ($E_{\gamma}$) may also escape depending upon the optical depth of the gas up to the domain edge. The optical depth is defined as,
 \begin{equation}
     \tau(r) = \int_{r}^{R_{e}} \rho(r^{\prime}) Y_{e}(r^{\prime}) \kappa_{\gamma} dr^{\prime}, 
 \end{equation}
where $R_{e}$ is the domain edge in the radial direction,  $\rho$ is the density, $Y_{e}$ is the electron fraction, and $\kappa_{\gamma}$ is the optical opacity. In our case, similar to \cite{stockinger2020}, we assume that Compton scattering is the dominant opacity source, with $\kappa_{\gamma} = 6.0 \times 10^{-2}$~cm$^{2}$~g$^{-1}$ \citep{Swartz_1995ApJ...446..766S}. The energy that is deposited locally is computed as $E_{\gamma} (1 - \exp[-\tau(r)])$.

\subsection{Grid Numerics} \label{sec:grid_numerics}
Our simulation grid very closely follows the FLASH setup in \citetalias{Sandoval_2021}. At the time of initialization in Flash-X, there are $2688 \times 192 \times 384$ cells in the $r$, $\theta$, and $\phi$ with $168 \times 24 \times 48$ blocks. The domain extends radially from 500~km to $1.51 \times 10^{9}$~km. The blocks are logarithmically spaced such that $\Delta r/r$ is nearly a constant, with values ranging from $5.8 \times 10^{-3}$ to $5.3 \times 10^{-3}$.  With a wedge of 5$^\circ$ around the $\theta$ boundary, the grid covers $0.0278\pi \leq \theta \leq 0.972\pi$ with $\delta\theta = 0.885^{\circ}$ and azimuthal angles $0 \leq \phi \leq 2\pi$ with $\delta\phi = 0.938^{\circ}$. For the decay phase of the simulation, the $\theta$ and $\phi$ blocks are reduced to $6 \times 12$ from $24 \times 48$ while keeping the same angular coverage and resolution. This remapping is performed due to the reduced decay network cost. 

\subsection{Progenitor model and Collapse simulation} \label{sec:prog_chimera}
Our simulation utilizes a 9.6-\msun\ zero-metalicity progenitor from A. Heger (private communication).
This progenitor has been used for 3D simulations by \cite{Melson2015a, Muller2019, stockinger2020, Sandoval_2021}, thus allowing us to compare the effects of additional physics on the evolution until shock-breakout. The progenitor star was evolved in KEPLER \citep{Weaver1978}, a 1D stellar evolution code, as an extension of the \citet{HeWo10} zero-metal set until  collapse, at which time it is mapped to \chimera.  \chimera\ \citep{Bruenn2020} is a neutrino radiation hydrodynamics code, where the collapse, formation of the bounce shock and finally the revival of the shock is simulated. Simulations in \chimera\ are carried out until the explosion energy has saturated and reached its asymptotic value. This particular \chimera\ model is a part of the D-series runs \citep{LeHiHa26}, and includes a 160-isotope nuclear reaction through XNet coupled with the radiation hydrodynamics. The coupled nuclear reaction network provides more detailed nucleosynthesic yields, particularly for neutron rich ejecta \citep{Harris_2017}.

We are using the \chimera\ model D9.6-sn160-3D, same as \citetalias{Sandoval_2021} for our principle simulation and the 1D and 2D equivalents for 1D and 2D runs. 
All the 1D and 2D simulations were mapped from \chimera\ to Flash-X at 0.650~s post-bounce. The explosion energy for the 1D \chimera\ (D9.6-sn160-1D) model was $4.3 \times 10^{49}$~erg at this point including on- and offgrid overburden and $1.9 \times 10^{50}$~erg for the 2D \chimera\ model (D9.6-sn160-2D).  The 3D model was mapped from the \chimera\ D9.6-sn160-3D model at 466.6~ms post-bounce with a total energy of $1.67 \times 10^{50}$~erg.
Since the simulation domain in \chimera\ only contains a small fraction of the entire progenitor, while mapping the profile to Flash-X, we map the \chimera\ model close to the edge of \chimera\ domain and add the progenitor profile until the edge of the progenitor. Beyond the progenitor, we map a steady stellar wind profile as described in Section~\ref{sec:csm_section}. As our simulation progresses, we continue to extend the domain with the steady wind CSM profile for density.

\section{1D and 2D evolution} \label{sec:1d_2d_results}
Ideally, we would run full 3D simulations with each incremental improvement in the included physics to understand the impact of that improvement, but that proves too computationally costly.
Instead, to understand the effects of different physics included in our simulations, a  series of 1D and 2D simulations were run for  the 9.6-\msun\ progenitor model.   The 1D and 2D simulations are evolved until $4 \times 10^{7}$~s and $3.024 \times 10^{7}$~s post-bounce, respectively.

\begin{figure*}
    \centering
    \includegraphics[width=\linewidth]{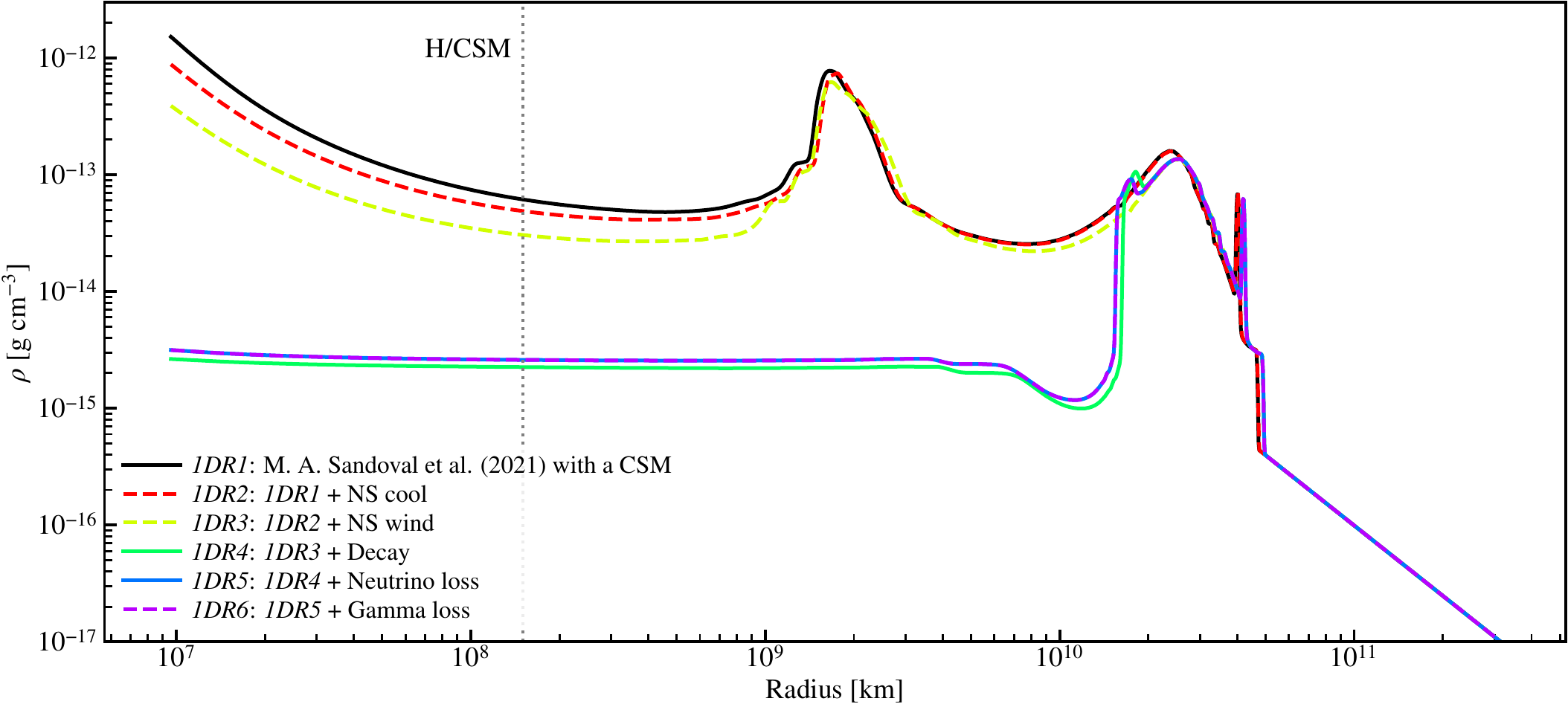}
    \caption{Density profile resulting from series of 1D runs at $t = 4 \times 10^{7}$~s, highlighting effects of different physics.  Vertical dashed gray line at $r = 1.5 \times 10^{8}$~km represents progenitor surface at time of mapping to Flash-X.}
    \label{fig:series_1d_dens}
\end{figure*}

\subsection{1D simulations}
In 1D, we ran a sequence of six simulations, labeled \runlabel{1}{DR}{1} through \runlabel{1}{DR}{6}, where each successive run includes one additional physics component relative to the previous one. The resulting density profile from these series of 1D runs at the end of the simulation is shown in Figure \ref{fig:series_1d_dens}.

\runlabel{1}{DR}{1} represents the 1D equivalent of the \citetalias{Sandoval_2021} setup, but with a CSM, which allows extension of our simulations much beyond shock-breakout. 
In this run, the full nuclear reaction network is turned on until 2.65~s post-bounce, after which the network is turned off and the simulation is purely hydrodynamical.  
At the time of mapping, the shock front is moving into the He layer from the C+O layer. 
As the simulation progresses, the shock front moves outward in both mass and radius, and interacts with the He/H interface at $\approx 1.4 \times 10^{7}$~km.  
The interaction of the shock with density and compositional gradients at this interface leads to the formation of a reverse shock, which moves inward in mass, engulfing the metal-rich ejecta, but outward in radius, reaching $\approx 1.7 \times 10^{9}$~km at the end of the simulation. 
As the shock enters the CSM, the shock-CSM interaction forms a forward shock, a reverse shock, and a contact discontinuity that separates the ejecta from the CSM.  The forward shock continues to progress through the CSM and is located at $\approx 5.0 \times 10^{10}$~km at the end of simulation. 
As the forward shock sweeps up CSM, the interaction shell decelerates. In the frame of the contact discontinuity, shocked ejecta from behind and shocked CSM from ahead both flow toward the contact, where the material accumulates because it cannot cross the interface. This compression produces the narrow density spike, which by the end of the simulation is located at $\approx 4.0 \times 10^{10}$~km. 
The broader hump immediately behind the spike corresponds to the bulk of the reverse-shocked ejecta, with a peak density at $\approx 3.5 \times 10^{10}$~km.

\runlabel{1}{DR}{2} includes NS cooling and accretion as described in section \ref{sec:pns_wind_cool}. As the NS cools by emitting neutrinos, the gravitational mass of the NS decreases, and hence the star becomes less bound.  
This has negligible effect on the shock and high velocity ejecta just behind it, but results in the  inner ejecta, inside of $2 \times 10^{9}$~km at the end of the simulation, being slightly more evacuated compared to \runlabel{1}{DR}{1}. 
\runlabel{1}{DR}{3} adds a NS wind to the physics included in \runlabel{1}{DR}{2}. 
A wind BC as described in section \ref{sec:pns_wind_cool} is applied from the inner boundary until 2.65~s post-bounce. 
This results in markedly lower densities in the innermost ejecta ($r < 10^{9}$~km) as the wind accelerates this slower moving material into the reverse shock peak, which is also pushed slightly outward.  
Effects of the wind are felt as far as $r \approx  5 \times 10^{10}$~km, with the high velocity ejecta and shock pushed slightly ahead compared with \runlabel{1}{DR}{2}. 
Of all of the physics added in this work, the wind has the largest effect on the shock's progress.
\runlabel{1}{DR}{4} includes a decay-only network which allows the radioactive isotopes to decay and to deposit energy locally. 
Because of the decay, a hot bubble forms and we observe the most significant effect on the density profile, lowering the density in the interior by more than an order of magnitude. 
The heating also strongly suppresses the progress of the reverse shock, which sits at $r \approx  1.5 \times 10^{10}$~km in this run, limiting the development of the innermost density peak seen in \runlabel{1}{DR}{1-3}.  
However, since the radioactive isotopes remain behind the shock front, the shock front profile is minimally affected, and thus similar to \runlabel{1}{DR}{3}. 
A loss of 20\% of decay energy in the form of neutrinos is allowed in \runlabel{1}{DR}{5}. 
Because of the energy loss, the bubble formed due to decay is modestly less effective, both at reducing the interior density and suppressing the reverse shock. 
In addition to neutrino losses, \runlabel{1}{DR}{6} also takes into account gamma energy losses that depend on optical depth. 
Since the ejecta is always behind the shock, the optical depth remains very high and hence the effect is inconsequential. 
Due to this, \runlabel{1}{DR}{6} lies on top of \runlabel{1}{DR}{5} in the density plot.

\subsection{2D simulations}

\begin{figure}
    \centering
    \includegraphics[width=\linewidth]{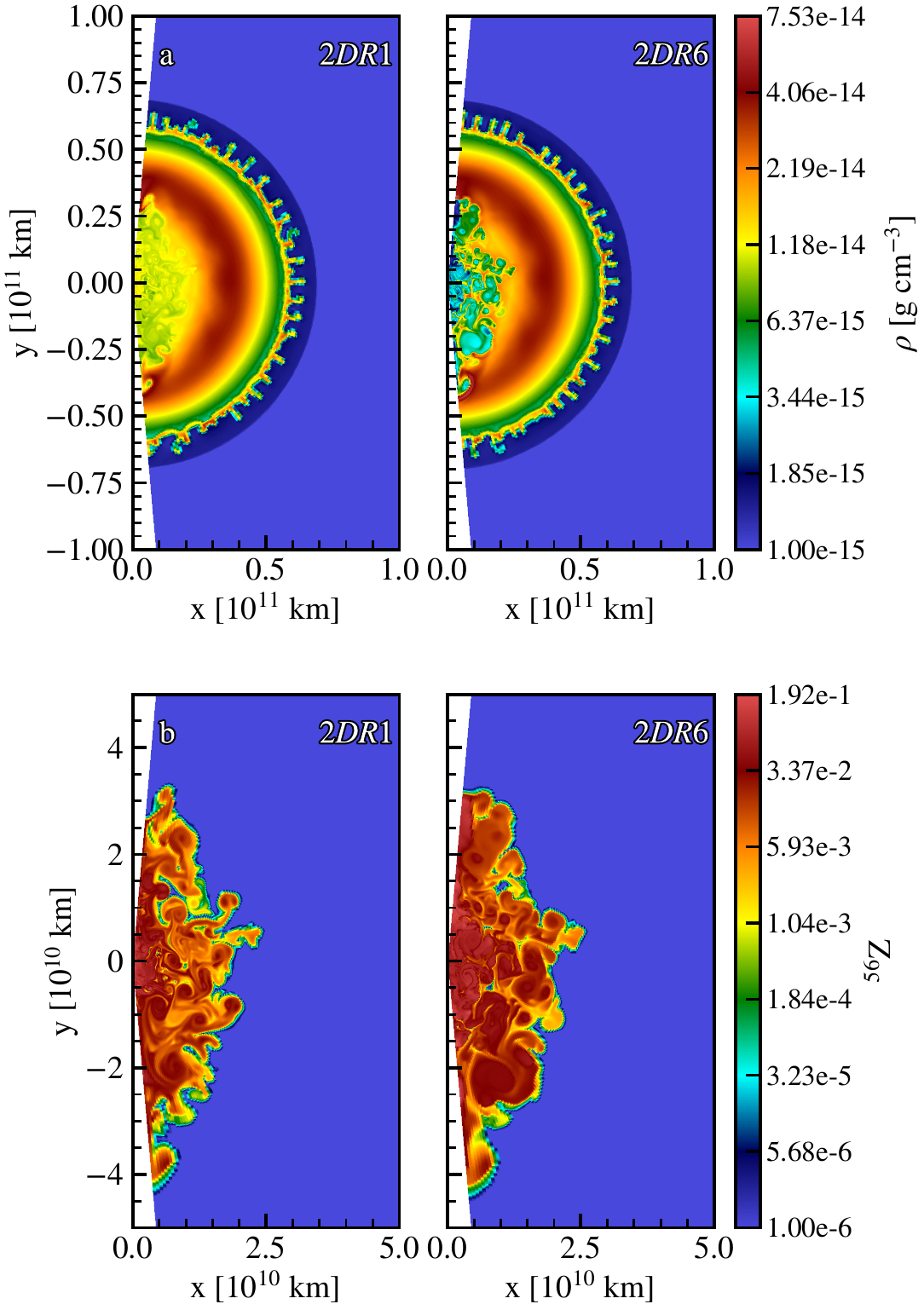}
    \caption{Comparison between \runlabel{2}{DR}{1} (left) and \runlabel{2}{DR}{6} (right) after 350 days. Top panel (a) shows the density profile, while the bottom panel (b) shows the mass fraction of \isotope{Z}{56} (defined as \isotope{Ni}{56}+\isotope{Co}{56}+\isotope{Fe}{56}). Note the (b) panel is zoomed in by a factor of 2.}
    \label{fig:2dsph_dens_z56}
\end{figure}

To examine the combined effect of the additional physics in 2D, we performed 2D counterparts of \runlabel{1}{DR}{1} and \runlabel{1}{DR}{6}, labeled \runlabel{2}{DR}{1} and \runlabel{2}{DR}{6}, respectively.
Because the 2D \chimera\ model has a higher explosion energy than its 1D counterpart, quantitative differences between models with the same physics are not solely the result of dimensionality in the Flash-X runs. 
For example, Figure \ref{fig:2dsph_dens_z56}(a), which shows the density profile for \runlabel{2}{DR}{1} and \runlabel{2}{DR}{6}, has the peak density at $r > 3 \times 10^{10}$~km even in \runlabel{2}{DR}{1}, which is more akin to \runlabel{1}{DR}{4-6}.
This is likely due to greater thermal and kinetic energy in the hot bubble, the result of convective neutrino heating in the 2D \chimera\ model. 
As in the case of 1D, the shock in \runlabel{2}{DR}{6} is slightly advanced by $3 \times 10^{8}$~km in radius over that of \runlabel{2}{DR}{1}.
In 1D, this was due to the NS wind, and the same is likely true here.
Note that the narrow high density shell ($r \approx 6 \times 10^{10}$~km in Figure \ref{fig:2dsph_dens_z56}(a)) just behind the shock is also slightly advanced in \runlabel{2}{DR}{6} as it was in \runlabel{1}{DR}{6}.
However, in 2D this shell serves as the base for a forest of RT plumes.
While the exact placement of the plumes is different for the 2 runs, as one would expect for a stochastic process, the typical height and spacing of the plumes is quite similar.

The impact of the radioactive decay, which separates \runlabel{1}{DR}{3} and \runlabel{1}{DR}{4} in Figure~\ref{fig:series_1d_dens}, is visible in Figure \ref{fig:2dsph_dens_z56}(a) as
well as  Figure \ref{fig:2dsph_dens_z56}(b), which displays profiles of the metal-rich ejecta \isotope{Z}{56} (defined as \isotope{Ni}{56}+\isotope{Co}{56}+\isotope{Fe}{56} mass fractions).
The effect is changed by localization of radioactive species.  
With the addition of decay heating, rather than the entire inner region reaching a uniform lower density as it did in 1D, in multi-D, pockets of metal-rich  material heat themselves, expanding to reach lower density while compressing metal-poor regions around them. 
This accentuates the density contrast between parcels in  \runlabel{2}{DR}{6} compared to \runlabel{2}{DR}{1} inside of $4 \times 10^{10}$~km (see Figure~\ref{fig:2dsph_dens_z56}(a)).
The interaction between the high pressure, radioactively heated matter and the reverse shock is also different between 1D and 2D, with the high-pressure/low-density parcels penetrating into the high density reverse shock region, rather than simply compressing it.
As a result, the metal-rich bubbles also extend as much as $5 \times 10^{9}$~km further in radius  in \runlabel{2}{DR}{6} compared to \runlabel{2}{DR}{1} (see Figure~\ref{fig:2dsph_dens_z56}(b)).
This is a considerably larger effect than the change in shock radius noted in the previous paragraph.

\begin{figure*}
    \centering
    \includegraphics[width=\textwidth]{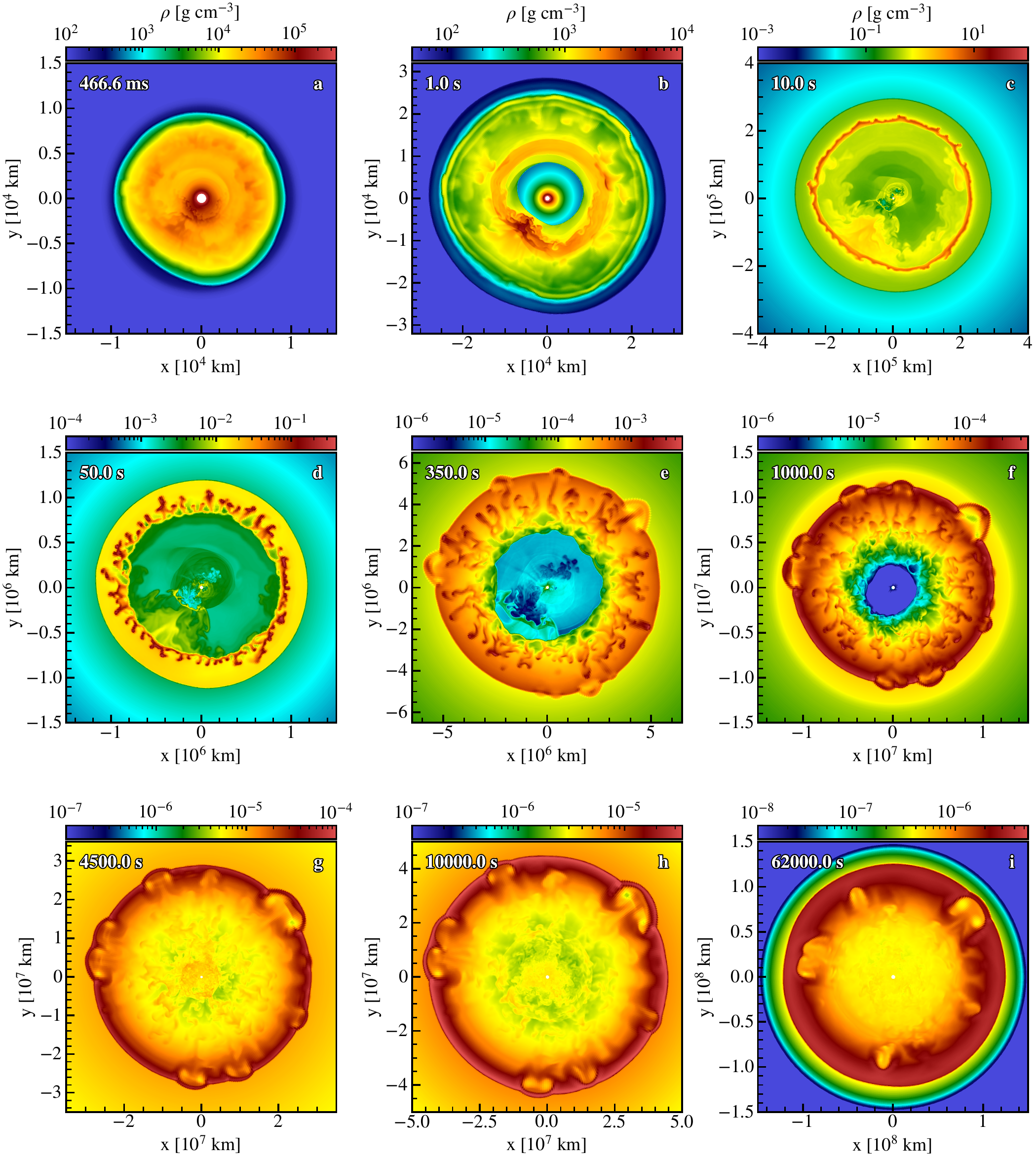}
    \caption{Density slice-plot of the $xy$ plane showing the evolution of plumes prior to shock-breakout.}
    \label{fig:dens_grid_prebreakout}
\end{figure*}

\begin{figure*}
    \centering
    \includegraphics[width=\textwidth]{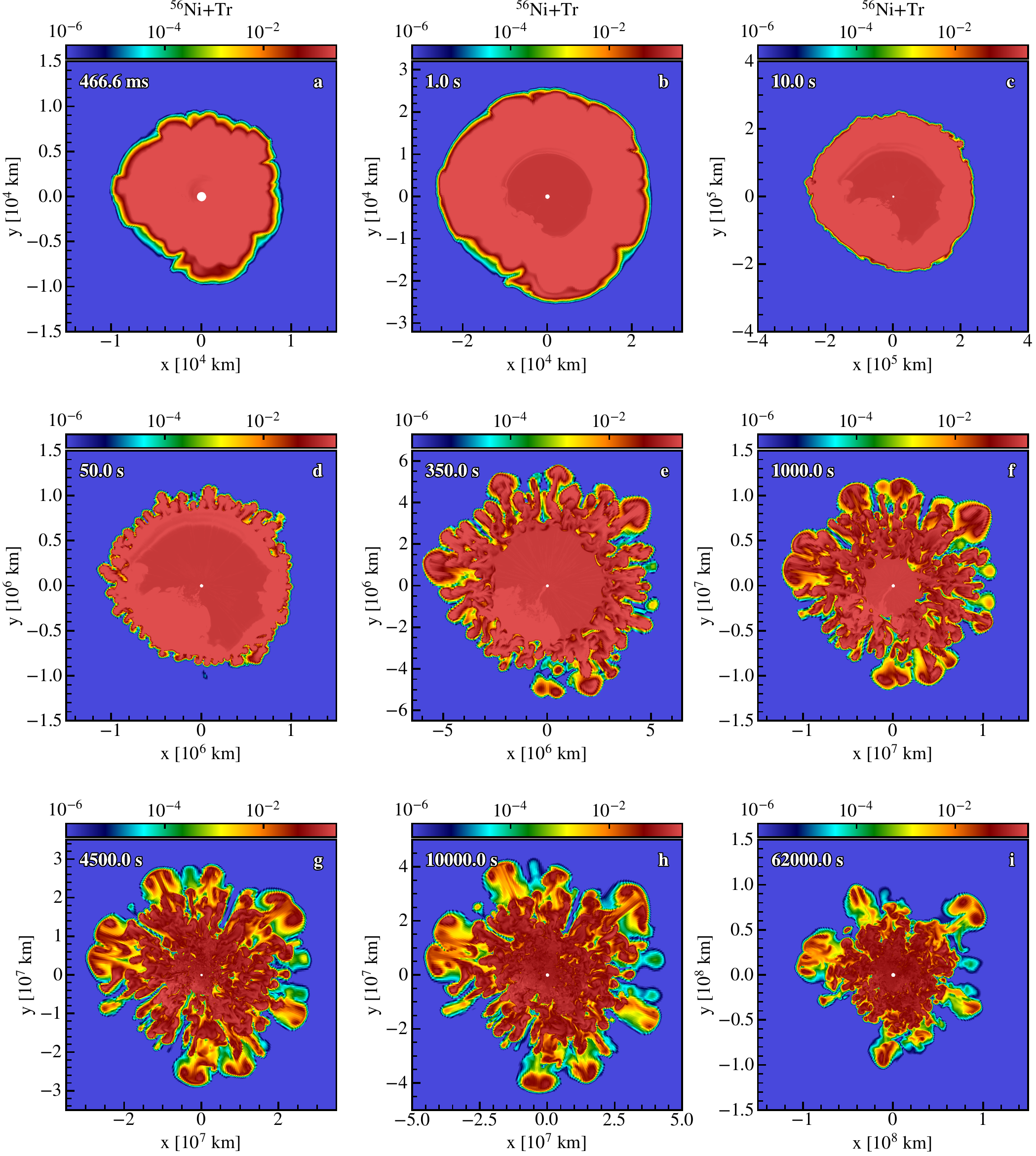}
    \caption{\nitr\ slice-plot of the $xy$ plane showing the compositional evolution of plumes prior shock-breakout.}
    \label{fig:ni56_trace_grid_prebreakout}
\end{figure*}

\section{Evolution in 3D}

The computational cost associated with a full 3D run allows us to run only the \runlabel{1}{DR}{6} equivalent 3D run. For comparison, we rely on the D9.6-3D3D model of \citetalias{Sandoval_2021}, which has the same physics as \runlabel{1}{DR}{1}, but without the CSM.  Prior to reaching the CSM, we can use this comparison to see the effects of the enhanced physics as described in section \ref{sec:methodology}.  
However, the use of different codes, Flash-X for this work verses FLASH for \citetalias{Sandoval_2021}, with different hydrodynamics schemes, adds an additional source of differences.  

Similar to the \citetalias{Sandoval_2021} D9.6-3D3D model, our 3D model is mapped at 466.6 ms post-bounce from \chimera\ to Flash-X with the inner $500$~km removed from the domain.
This includes the NS, which is treated as a point-mass at the origin. 
The diagnostic explosion energy at the end of the \chimera\ 3D run is $1.68 \times 10^{50}$~erg. 
As the 3D run progresses, we drop the blocks whose radii are less than 2$\%$ of the minimum main shock radius, similar to \cite{stockinger2020}. 
For comparison with the more limited chemical composition evolved in \cite{stockinger2020}, like  \citetalias{Sandoval_2021}, we define neutron-rich iron group as all species in our network falling in the range of \isotope{Cr}{49}--\isotope{Ni}{64}, excluding \isotope{Fe}{52} and \isotope{Ni}{56}. Metal-rich bullets, labeled as \nitr, are defined by the sum of the mass fractions of \isotope{Ni}{56} and neutron-rich iron group. 
Data from select snapshots beginning with shock break out are available in Flash checkpoint format at the OLCF Constellation repository: \dataset[doi: 10.13139/OLCF/3367469]{\doi{10.13139/OLCF/3367469}}.

\subsection{Before shock-breakout}
The overall dynamics until shock-breakout in our simulation closely match those reported in \citetalias{Sandoval_2021}. 
We therefore focus here on a few key stages of the evolution, illustrated in Figure \ref{fig:dens_grid_prebreakout}, which displays the density and Figure \ref{fig:ni56_trace_grid_prebreakout}, which shows the mass fraction of \nitr.
By the mapping point from \chimera\ to Flash-X, 466.6~ms post-bounce, the shock has already traversed the (C+O)/He interface, as shown in Figure~\ref{fig:dens_grid_prebreakout}(a), and has reached a mean radius of $\approx 1 \times 10^{4}$~km. 
The NS wind stalls accretion and drives the innermost material outward. 
By 1.0~s post-bounce, the wind has injected $1.2 \times 10^{-3}$~\msun\ into the domain, carved out a low-density cavity around the NS, and maintained a dense layer behind the shock by suppressing continued accretion (Figure~\ref{fig:dens_grid_prebreakout}(b)). 
Despite a spherically symmetric wind, the asymmetries left behind by the explosion in the region between the NS and the shock front result in an asymmetrical cavity.
At this stage, the NS wind is turned off and the nuclear reaction network is reduced to decay only. 
Even after the wind is shut off, the momentum it has deposited continues to drive the cavity outward, although accretion persists near the inner boundary.
By 2.0~s, the cavity expansion has stalled and begun to reverse as accretion becomes stronger. 
The shock is then fully in the He layer and, because of the compositional gradient, seeds for RT plumes have formed in regions behind the shock. By 10.0~s, continued accretion to the NS  has erased any resemblance of the cavity near the inner boundary. 
At this stage, the shock is at $\approx 3 \times 10^{5}$~km and has slowed down by $\approx 1.1 \times 10^{4}$~\kmps\ to $2.5 \times 10^{4}$~\kmps\ due to the positive $\rho r^{3}$ gradient. 
The RT unstable ejecta formed in the post-shock region can be seen in Figures \ref{fig:dens_grid_prebreakout}(c), \ref{fig:ni56_trace_grid_prebreakout}(c). 
The metal-rich plumes continue to keep up with the shock front and avoid being crushed by the reverse shock. 

\begin{figure}
    \centering
    \includegraphics[width=\linewidth]{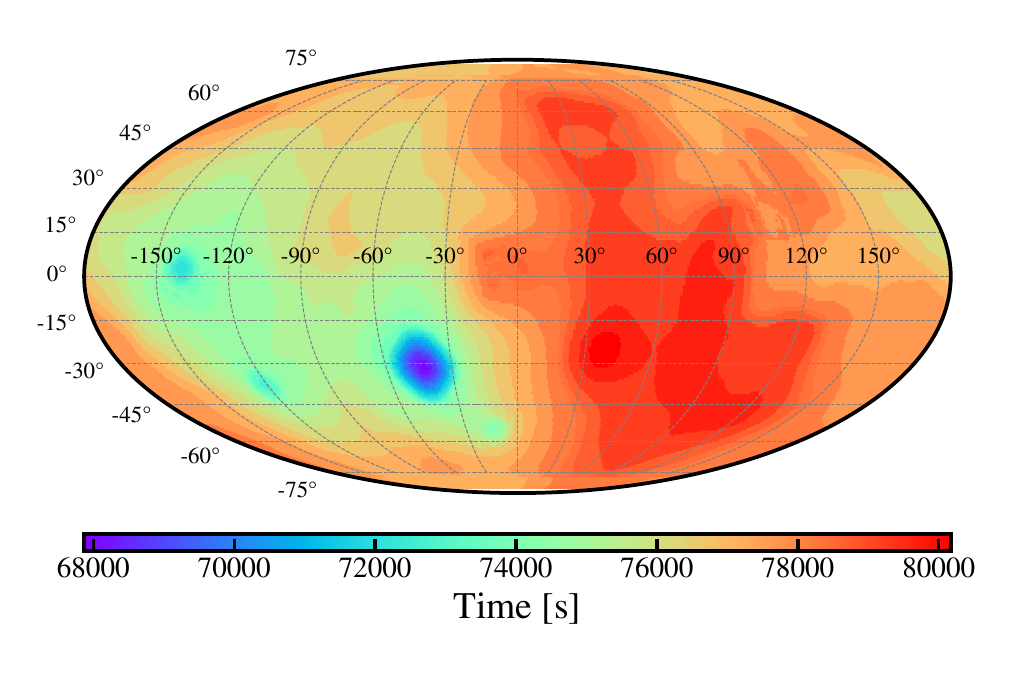}
    \caption{Mollweide projection of shock breakout of progenitor surface in each direction.}
    \label{fig:shock_breakout_span}
\end{figure}

\begin{figure}
    \centering
    \includegraphics[width=\linewidth]{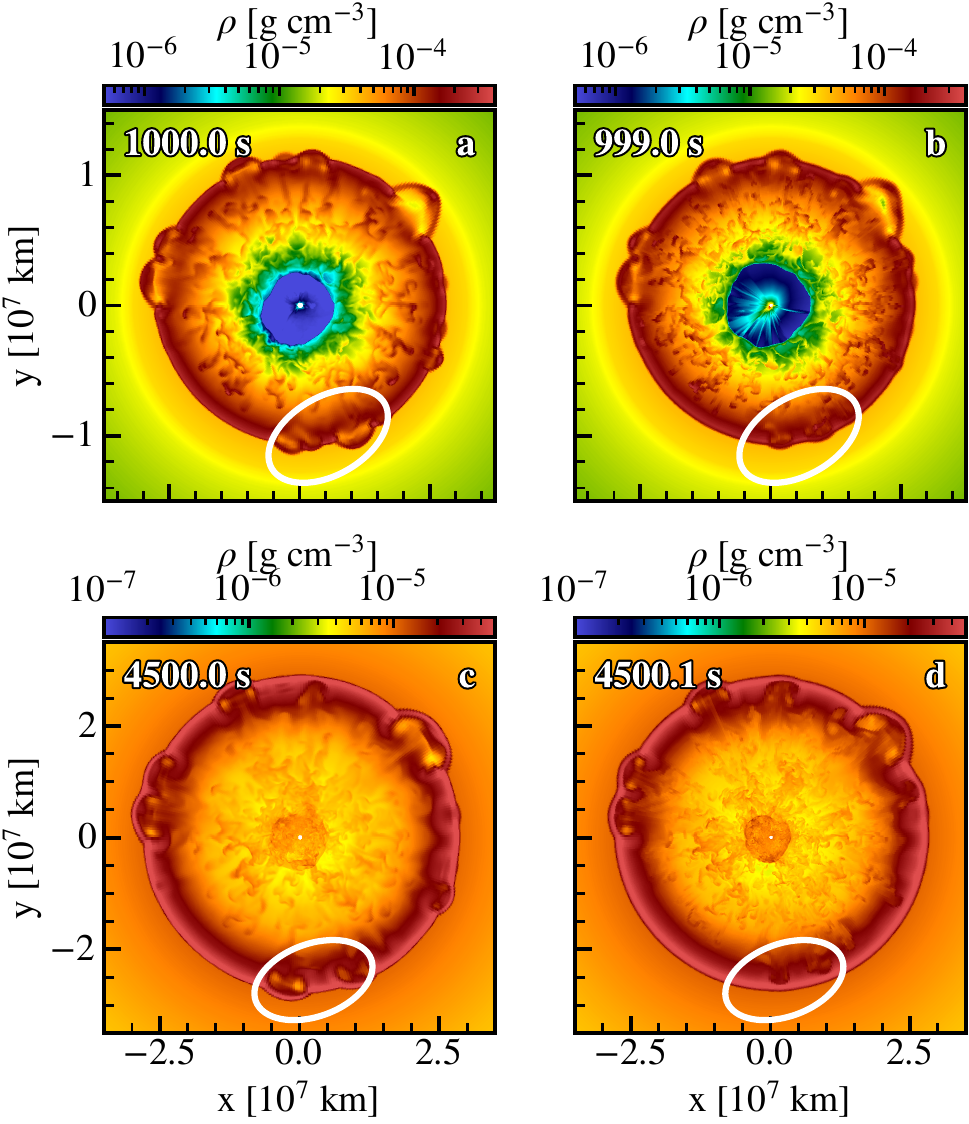}
    \caption{Density plot in the $xy$-plane at 1000~s and 4500~s comparing this work, using Flash-X (left), and \citetalias{Sandoval_2021} D9.6-3D3D model, using FLASH (right). White ellipse shows region in Flash-X run that forms additional plumes.}
    \label{fig:plume_growth_flash4_flashx}
\end{figure}

\begin{figure*}
    \centering
    \includegraphics[width=\linewidth]{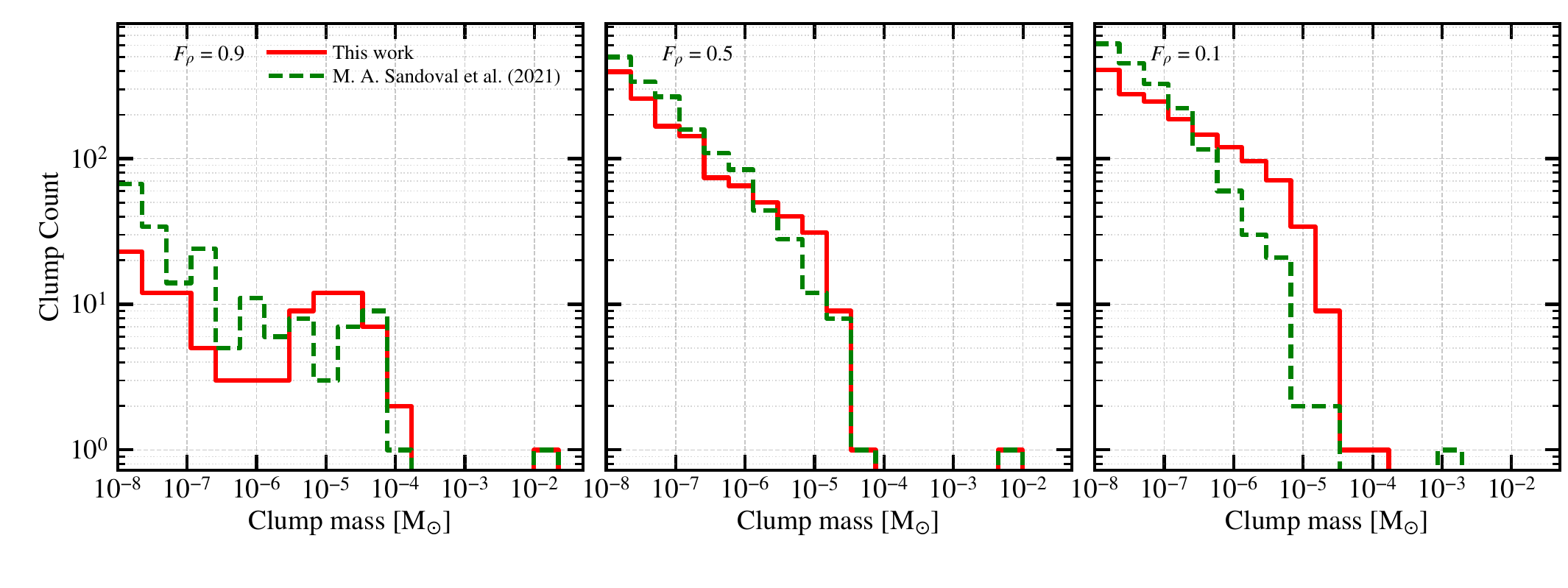}
    \caption{Clump count comparison between this run and \citetalias{Sandoval_2021} for different $F_{\rho}$ at 62000~s.}
    \label{fig:clump_count}
\end{figure*}

The RT features continue to grow as the shock moves through the He shell, forming more pronounced features at 50.0~s (Figure \ref{fig:dens_grid_prebreakout}(d)). 
These features have metal-rich bullets, as shown in Figure \ref{fig:ni56_trace_grid_prebreakout}(d). 
By 350~s, the RT plumes along with the metal-rich ejecta have become larger and a few of the plumes have caught up with and even penetrated the main shock front, which can be seen in Figure~\ref{fig:dens_grid_prebreakout}(e) as the density gradient present (yellow to green transition stripe) at $\approx 5 \times 10^{6}$~km. 
The reverse shock that formed when the shock front propagated through the (C+O)/He interface continues to move towards the inner boundary (blue region at $\approx 2.5 \times 10^{6}$~km). At 1000~s, the shock is close to the He/H interface and about to move into the H shell. 
In Figure \ref{fig:dens_grid_prebreakout}(f), we can see the density gradient present at the interface at $\approx 1.4 \times 10^{7}$~km. 
As the shock moves through the interface, the density gradient results in the formation of a weak pressure wave which moves inward in mass and radius. As the metal-rich ejecta moves into the H layer, it is coated with helium.

The pressure wave that developed as the shock moved into the H layer eventually forms into a second reverse shock and interacts with the inner radial boundary at $\approx 20000$~s. 
The shock, along with the ejecta, continues to propagate through the hydrogen layer with the metal rich RT plumes getting modified and growing in size faster than the overall expansion of the metal rich region (Figures \ref{fig:dens_grid_prebreakout} (g,h,i)).  
The shock front first comes into contact with the progenitor edge at $\approx 68000$~s, starting the shock breakout process (see Figure \ref{fig:shock_breakout_span}).

Compared to \citetalias{Sandoval_2021}, the inclusion of additional physics in our simulation alters both the morphology and the number of enriched ejecta plumes. In particular, the symmetric NS wind does not merely halt accretion, it also compresses the ejecta toward the forward shock. 
This supports the development of RT plumes along preferred directions that can keep pace with the shock and, crucially, avoid being shredded by the reverse shock. 
As a result, these plumes continue to grow and become increasingly prominent as the ejecta moves from the C+O layer into the He layer. 
This is shown in Figure \ref{fig:plume_growth_flash4_flashx}, which directly compares the \citetalias{Sandoval_2021} D9.6-3D3D run to our 3D model.  The white ellipse in the southern hemisphere highlights a region where additional plumes form in our enhanced physics simulation.

While the clumpiness of the metal-rich ejecta is obvious in Figure \ref{fig:ni56_trace_grid_prebreakout}, a more quantitative method to identify clumps, in a way that is not sensitive to an arbitrary absolute cutoff, is desirable.
We therefore define the clump-forming region using the \nitr\ density $\rho_X \equiv \rho \sum_i X_i$, where the sum runs over \nitr\ nuclei. 
Following \cite{gabler_2020}, we determine the threshold $\rho_{X,\min}$ by requiring that the subset of cells with $\rho_X \ge \rho_{X,\min}$ contains a fixed fraction $F_\rho$ of the total \nitr\ mass in the domain. 
In this method, $F_{\rho}$ specifies how much of the densest \nitr\ rich material is treated as a part of the clump forming ejecta. 
Once $ \rho_{X,\min}$ is determined, connected neighboring cells above this threshold are grouped into structures, which are counted as individual clumps or plumes. 
The lower value of $F_{\rho}$ selects only the densest parts of the metal rich ejecta, while higher value includes more diffuse material and can connect several dense regions into a larger structure.
With this approach, we find that our run, to about the same point in simulation time, produces fewer small plumes and more large plumes in general than \citetalias{Sandoval_2021} (Figure \ref{fig:clump_count}), particularly for $F_\rho = 0.1$. 
This is consistent with decay heating inflating the plumes and promoting their merger, leaving us with fewer but larger structures. 
However, since our model also includes additional physics effects (in particular the NS wind) and uses a different hydrodynamic scheme, it is not possible to establish precisely which change is driving this difference. 
Nevertheless, the overall clump statistics and morphology are consistent with what is expected from these added effects.

\begin{figure*}
    \centering
    \includegraphics[width=\linewidth]{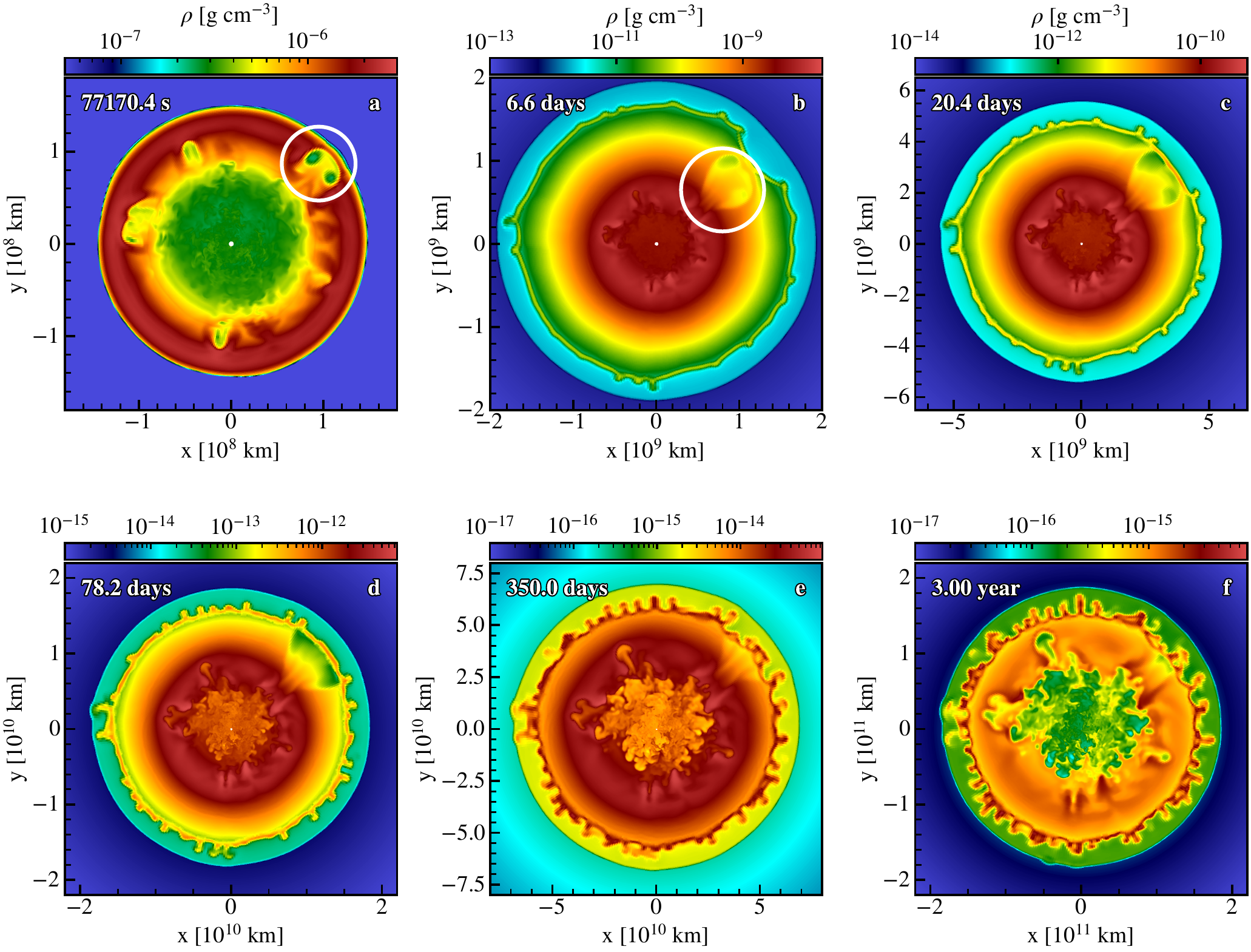}
    \caption{$xy$-plane density slice showing the evolution CSM after shock breaks out of progenitor.}
    \label{fig:dens_grid_post_breakout}
\end{figure*}

\begin{figure}
    \centering
    \includegraphics[width=\linewidth]{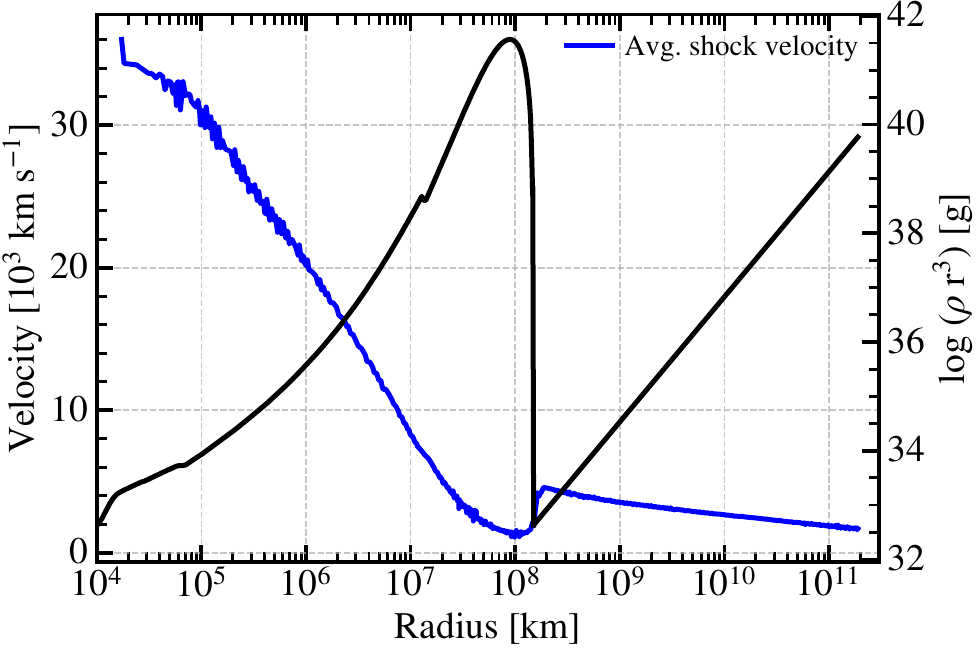}
    \caption{Average shock velocity (blue) of shock propagating through the progenitor and (black) progenitor $\rho r^{3}$ profile.}
    \label{fig:avg_shock_velx}
\end{figure}

\begin{figure*}
    \centering
    \includegraphics[width=\linewidth]{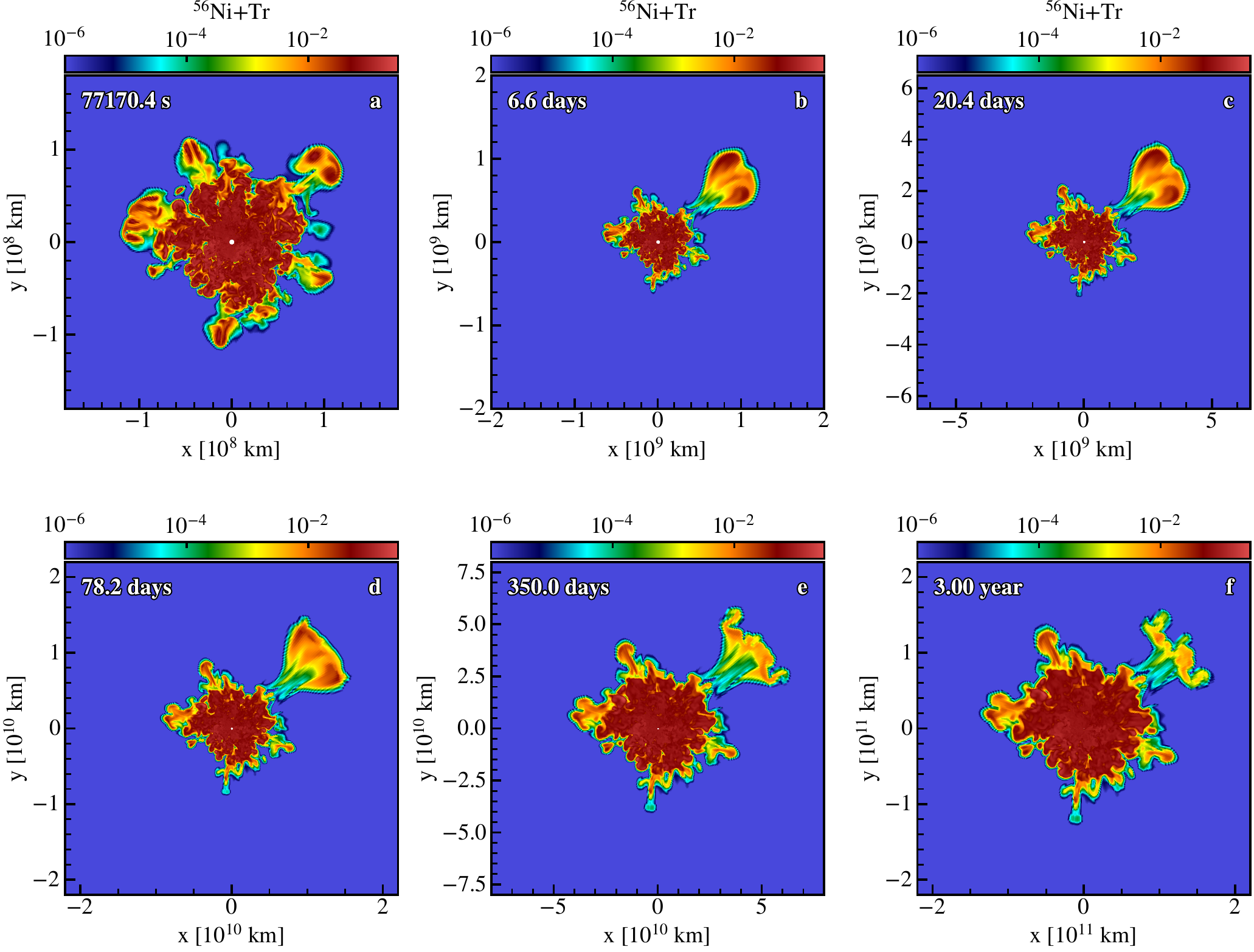}
    \caption{$xy$-plane slice showing the evolution of \nitr\ mass fraction in CSM. Note change in plot limits as the ejecta expands into CSM.}
    \label{fig:ni56_grid_post_breakout}
\end{figure*}

\subsection{Beyond shock-breakout}

In our simulation, shock breakout is not a single instant but an extended process that varies with latitude and longitude across the stellar surface, beginning at $\approx 68000$~s and lasting until $\approx 80000$~s, as shown in Figure~\ref{fig:shock_breakout_span}. 
This spread in breakout times reflects the aspherical shock front, with directions where elongated plumes have pushed the shock outward reaching the progenitor surface earlier than directions where the shock remains less extended. 
The average shock radius becomes larger than the progenitor edge at $\approx 77000$~s. 
At shock breakout, the large plumes, shown as the low-density green regions in Figure~\ref{fig:dens_grid_post_breakout}(a), have comparable radial extents, except for the plume highlighted by the white circle, which lies closer to the shock at $\approx 1.5 \times 10^{8}$~km and has a more coherent, nearly spherical head.

Once the shock emerges from the progenitor, it accelerates through the steep density drop at the stellar edge and propagates into the CSM, as shown by the sharp increase in shock velocity, from 2000~\kmps\ to more than 4500~\kmps, near the progenitor edge in Figure~\ref{fig:avg_shock_velx}. 
The sharp density gradient at the progenitor edge also leads to the formation of a third reverse shock. 
Although this reverse shock moves inward into freely expanding ejecta in mass coordinate, it propagates outward in radius.
The ejecta follows the shock into the CSM but cannot keep up with the rapidly expanding shock front. 
As the simulation progresses, the forward shock, the detached reverse shock, and the expanding ejecta move farther into the CSM.

The leading nearly spherical  plume experiences less deceleration than the other large plumes as its more coherent, compact structure is less susceptible to stripping and deformation by the material ahead of it as it expands through the CSM. 
As a result, it maintains its outward motion more effectively and approaches the reverse shock earlier. 
By 6.6~days, this plume, highlighted by the white circle in Figure~\ref{fig:dens_grid_post_breakout}(b), has moved well ahead of the other large plume structures, visible in the red ejecta region. 
This outsized plume is particularly notable when focusing on metal-rich regions (Figure ~\ref{fig:ni56_grid_post_breakout}(b)).
At this time, the separation between the forward shock, visible as a density gradient at $\approx 1.9 \times 10^{9}$~km, and the reverse shock, visible as a thin high-density near circular band at $\approx 1.6 \times 10^{9}$~km, is very prominent. 
By 20~days, the leading ejecta begins to catch up to the third reverse shock, and the largest plume starts to deform, as shown in Figure~\ref{fig:dens_grid_post_breakout}(c). 

\begin{figure*}
    \centering
    \includegraphics[width=\linewidth]{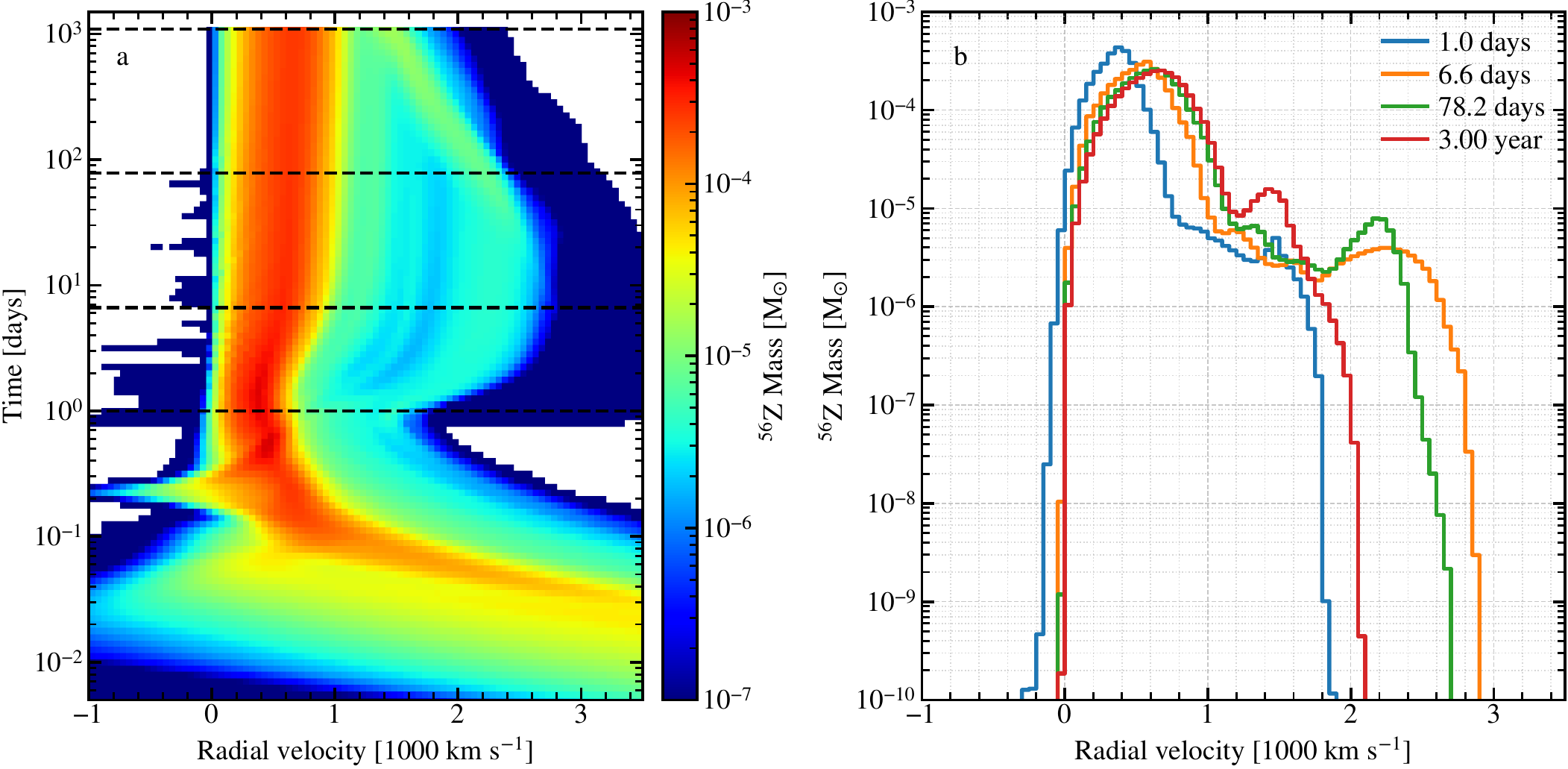}
    \caption{\isotope{Z}{56} (\isotope{Ni}{56}+\isotope{Co}{56}+\isotope{Fe}{56}) evolution in radial velocity space (90 bins of width $50$~\kmps\ ) for (a) entire evolution; (b) histograms at four times marked with horizontal dashed lines in left plot.}
    \label{fig:a56_velx_evolution}
\end{figure*}

As time progresses, other large-scale plumes also reach the reverse shock and begin to deform. 
By 78.2~days, the head of the previously nearly spherical metal-rich plume is no longer spherical and is being distorted by its interaction with the reverse shock, as seen in Figures~\ref{fig:dens_grid_post_breakout}(d) and~\ref{fig:ni56_grid_post_breakout}(d). 
The interaction of these plumes with the reverse shock decelerates them and seeds smaller-scale structure along the interface. 
In this sense, the large-scale asymmetry present at breakout is not erased immediately, but is gradually transformed into a more fragmented, small-scale structure by the ejecta-reverse shock interaction.
During this time, the inner plumes continue to expand and move closer to the reverse shock.

Most of the new small-scale plumes produced by the ejecta-reverse shock interaction in the CSM are relatively metal-poor, although some metal-rich structures also form near the base of the expanding ejecta, as seen by comparing Figures~\ref{fig:dens_grid_post_breakout}(e,f) and~\ref{fig:ni56_grid_post_breakout}(e,f). 
This comparison shows that not every density plume corresponds to a metal-rich plume: some of the late-time developing, finger-like structures seen in the density field are produced mainly by the hydrodynamic interaction with the reverse shock, while the metal-rich material remains more concentrated in the inner ejecta and in a smaller number of stronger plumes.

After the outer ejecta structures encounter the reverse shock, their subsequent evolution becomes roughly self-similar, with continued expansion, deceleration, and small-scale fragmentation at the interface. 
By 3~year, the initially coherent spherical plume has been almost completely shredded by the reverse shock (see Figures~\ref{fig:dens_grid_post_breakout}(f)), although it is still visible as enhanced metals in Figure~\ref{fig:ni56_grid_post_breakout}(f). 
The inner plumes that initially lagged behind have now expanded outward and approached the reverse shock, although some of them have not yet fully interacted with it. 
At these late times, additional non-thermal physics, such as cosmic-ray acceleration, cosmic-ray feedback, and radiative losses (other than escape of $\gamma$-rays), which are not included in our simulation, may become important in further shaping the ejecta dynamics and morphology. 
We therefore stop the simulation at this stage.

The evolution of the \isotope{Z}{56} (defined as \isotope{Ni}{56}+\isotope{Co}{56}+\isotope{Fe}{56}) mass in radial velocity space over time is shown in Figure~\ref{fig:a56_velx_evolution}(a). 
The initial hours of the explosion see the high velocity of the metal-rich core of the ejecta (represented here by the red and orange colors) experience strong deceleration as they drive the stellar envelope outward, reaching a minimum near shock breakout.
Ejecting the stellar envelope has a similar effect on the metal-rich plumes that have developed during the explosion (represented here by the blue regions of Figure~\ref{fig:a56_velx_evolution}(a)).
A new round of plumes are born as a result of shock breakout, reaching maximum  velocities of 3000~\kmps\ (Figure~\ref{fig:a56_velx_evolution}(b)), within a few days after breakout. 
At the same time, the metal-rich core material is also accelerating, reaching more than $\sim 1000$~\kmps, while the peak of the distribution accelerates from 350~\kmps\ bin at 1 days to 550~\kmps\ bin at 6 days and continues to get broader. Here, binned velocities are labeled by the left edge of each velocity bin.

\begin{figure*}
    \centering
    \includegraphics[width=\linewidth]{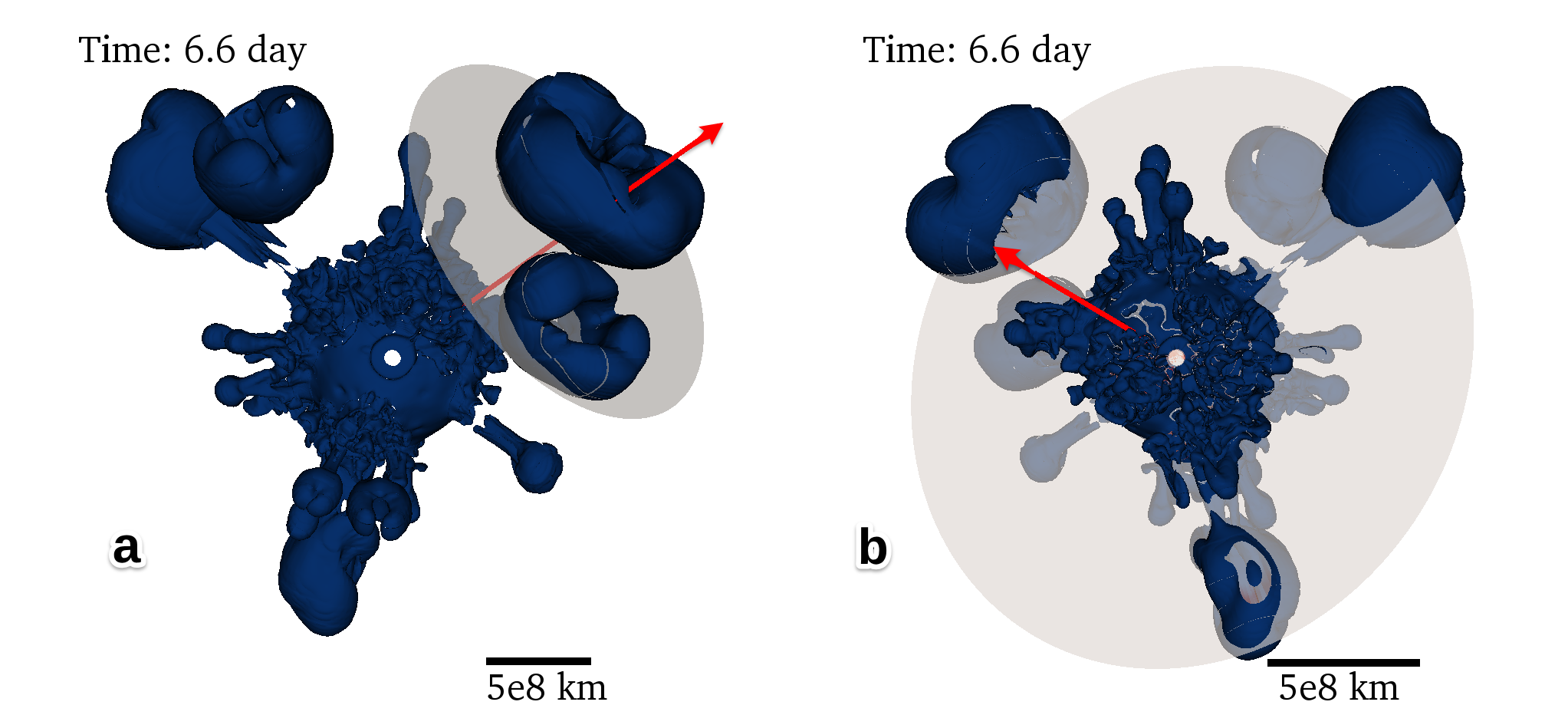}
    \caption{\isotope{He}{4} iso-surface (blue) showing the plumes which contain the metal-rich ejecta, along with the planes that define the two line of sight (LOS) directions, represented by red arrow, at 6.6~days post-bounce for (a)  plane with normal (\textit{LOS-max}) aligned with propagation of two of the largest plumes; and (b) plane with normal (\textit{LOS-min}) transverse to propagation direction of three largest plumes.}
    \label{fig:merged_los_plumes}
\end{figure*}

After 20 days, the \isotope{Z}{56} species embedded in the plumes are progressively decelerated by the reverse shock, with the peak in the distribution due to the plumes decelerating from 2200~\kmps\ bin at 6 days (orange line in Figure~\ref{fig:a56_velx_evolution}(b)) to 1400~\kmps\ bin at 3 years (red line in  Figure~\ref{fig:a56_velx_evolution}(b)).
This occurs gradually at first but more strongly after 80 days. 
In contrast, over the same time, the metal-rich core is slightly accelerating and becoming broader, with the peak of the distribution reaching 650~\kmps\ bin after 3 years, likely due to radioactive heating. 

\subsection{Observable Consequences}

To understand how the overall morphology of the metal-rich ejecta and especially these changes in radial velocity structure over time might manifest in observations, we construct two representative line of sight (LOS) directions sampling the large scale plume geometry, as shown in Figure~\ref{fig:merged_los_plumes}.
The first LOS is chosen to be approximately aligned with two of the large plumes, so that the plume material has a larger velocity component along the observer's direction. 
We refer to this direction as \textit{LOS-max}.
The second LOS is defined by the normal to the plane containing the three largest plumes. 
Since the dominant plumes lie mostly within this plane, this viewing direction samples the plume motion mostly perpendicular to their main axes and therefore gives the smallest projected LOS velocity spread, but the largest transverse velocity spread. 
We refer to this direction as \textit{LOS-min}.
These two LOS choices bracket the range of possible velocity-space distributions produced by viewing the same asymmetric ejecta morphology from different directions, highlighting the impact of viewing direction on this model.

\begin{figure*}
    \centering
    \includegraphics[width=\linewidth]{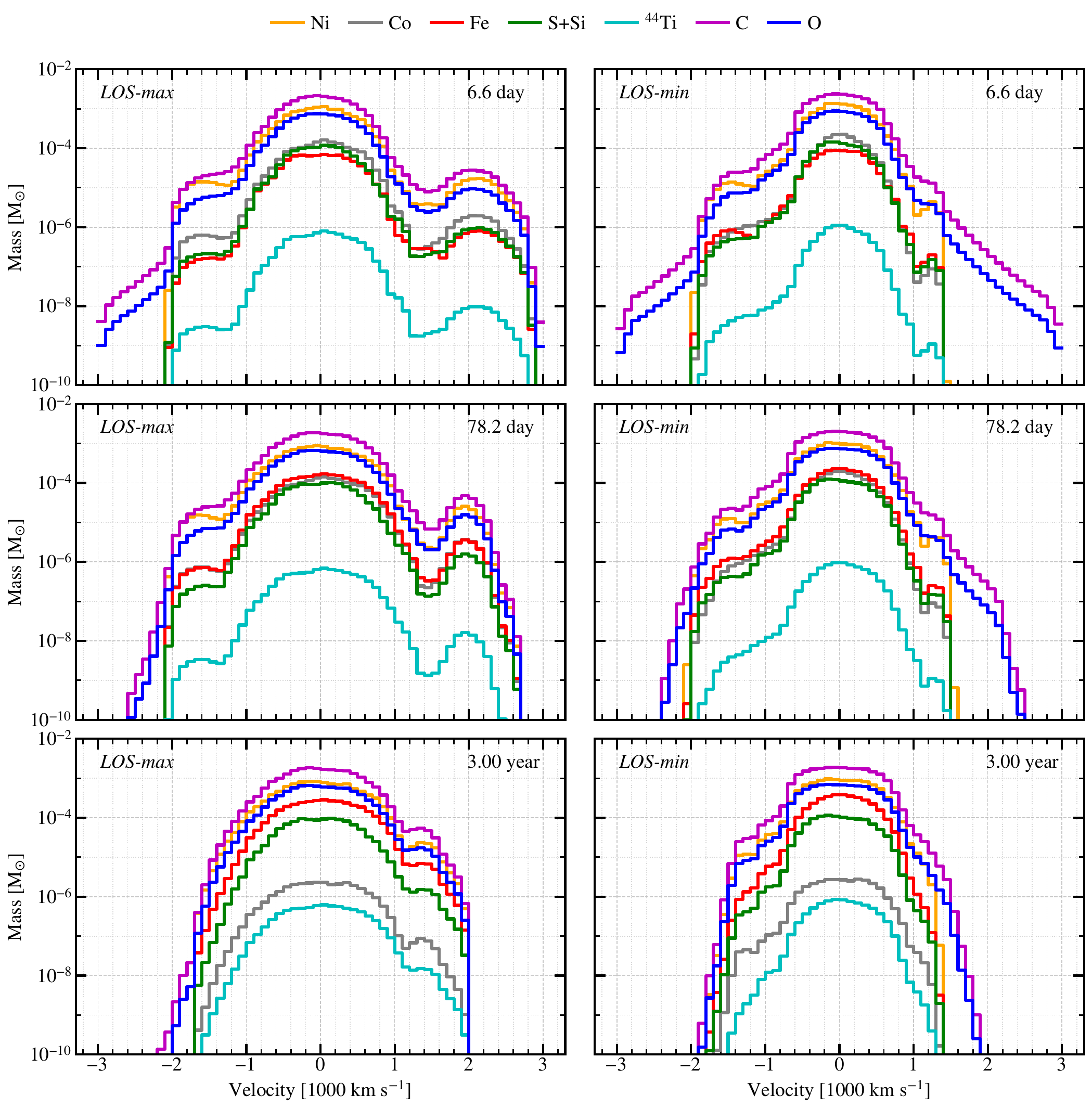}
    \caption{Line of sight velocity distributions (60 bins of width 100~\kmps\ ) for Ni, Co, Fe, Si+S, O, C, \isotope{Ti}{44}, along the \textit{LOS-max} (left) and \textit{LOS-min} (right) at 6.6 days (top row), 78.2 days (middle row), and 3 years (bottom row).} 
    \label{fig:grid_element_vlos_dist}
\end{figure*}

The ejecta distribution in LOS velocity space along the two representative viewing directions are shown in Figure~\ref{fig:grid_element_vlos_dist}. 
The distributions show a clear dependence on viewing angle, especially for $|\vlos| > 1000~\kmps$. 
At 6.6~days post-bounce, the \textit{LOS-max} direction exhibits three main components: a central component near zero, spanning  $\vlos \approx \left[-1000, 1000\right]~\kmps$, associated with the slower-moving inner, metal-rich ejecta, a strong positive-velocity component centered at $\vlos \approx +2100$~\kmps, and a weaker negative-velocity component extending to $\vlos \approx -1600$~\kmps. 
The high positive-velocity component is produced by the large plume-aligned structure in the \textit{LOS-max} direction, where the motion of the dominant plumes has a large projection along the line of sight. 
The negative-velocity component is produced by a large plume that is only partially aligned with the negative \textit{LOS-max} direction.
These plumes contain metal-rich ejecta and also sweep up C+O-rich progenitor material as they move outward. 
Because this material is pushed to large radii, it is further accelerated when the shock encounters the steep density gradient near the outer edge of the progenitor.

The \textit{LOS-min} direction also shows a similar multi-component structure at the same time, but the high-velocity plume component is significantly suppressed. 
In this direction, the positive-velocity feature is shifted inward to $\vlos \approx +1200$~\kmps, and the distribution is more centrally concentrated.
This occurs because \textit{LOS-min} is approximately normal to the plane containing the three largest plumes, so much of the plume motion lies closer to the plane than along the line of sight. 
As a result, the same 3D plume structure produces a smaller projected velocity spread in \textit{LOS-min} than in \textit{LOS-max}.

Along both LOS, there is indication of high velocity C- and O-rich material.
This is the result of dredge up from the helium burning shell into the envelope during the star's life.  
While this progenitor is initially metal free, the combination of helium burning and convection introduce small amounts of metals to the envelope of the star.
\isotope{C}{12} is most prominent, with a mass fraction of $4 \times 10^{-4}$ just above the helium burning layer ($r  = 1.45 \times 10^7$ km), declining steadily to a value of $1 \times 10^{-5}$ near the stellar surface ($r  = 1.47 \times 10^8$ km) before plummeting nearly to 0 in the next $10^6$ km, leaving a metal free layer 2 million km thick at the surface. 
Other isotopes follow the same pattern, notably \isotope{O}{16} (a factor of 4 smaller), \isotope{N}{14} ($850\times$ smaller), \isotope{Ne}{20} ($180\times$ smaller) and \isotope{Mg}{24} ($2000\times$ smaller).
There are even traces of silicon (6 orders of magnitude less than \isotope{C}{12}), iron and nickel (12 order of magnitude less) and s-process species (11 orders of magnitude less.)

This spherical component is most clearly visible on the negative-velocity side of the \textit{LOS-max} distribution, where C and O extend at 6.6 days smoothly from velocities of approximately $-2500$ to $-3000$~\kmps\ with low ejecta mass. 
However, on the positive-velocity side of \textit{LOS-max}, this underlying C+O distribution is largely masked by the plume-aligned component. 
Thus, the \textit{LOS-max} profile can be interpreted as the combination of a broad, nearly spherical C+O component and an additional localized enhancement from plume-entrained material. 

In contrast, \textit{LOS-min} more directly shows the underlying, centrally peaked C+O distribution and the spherical nature of the background C+O distribution accelerated to high positive and negative velocities by the steep density cliff at the progenitor edge. 
This is due to the dominant plume motion being in direction transverse to the line of sight.

The evolution beyond 6.6 days shows the effect of deceleration by the reverse shock. In the \textit{LOS-max} direction, the positive velocity peak distribution becomes more narrow, with the outer edge shifting inwards from $\approx +2900~\kmps$ at 6.6 days to $\approx +2700$~\kmps\ by 78.2 days, indicating that the fastest plume-aligned material has been slowed by the reverse shock. 
The outer C+O material, extending to high negative velocity, is slowed down more dramatically, from approximately $-3000$~\kmps\ to $-2600$~\kmps. 
During the same time, the width of the central metal-rich ejecta increases slightly.
Note also the slight increase in Fe, surpassing Si+S, and decline in Co, due to the decay chain $\isotope{Ni}{56} \rightarrow \isotope{Co}{56} \rightarrow \isotope{Fe}{56}$.
However, stable (neutron-rich) Ni remains much more abundant. 
By 3.0 years, the metal-rich ejecta distribution in the \textit{LOS-max} direction is more symmetrical, extending from $-1600$~\kmps\ to +2000~\kmps, with only low mass C+O material extending beyond this to more negative velocity. 

The \textit{LOS-min} case evolves differently in velocity, although the effects of the $\isotope{Ni}{56} \rightarrow \isotope{Co}{56} \rightarrow \isotope{Fe}{56}$ decay chain are also visible. 
The lack of large plumes along the positive LOS direction limits the effects of plumes encountering the reverse shock, resulting in nearly the same metal-rich ejecta distribution along this LOS at 78.2 days compared to 6.6 days. 
Only the outer C+O components gets decelerated by the reverse shock, contracting from $\left[-3000, 3000\right]~\kmps$ to $\left[-2400, 2600\right]~\kmps$. 
By 3.0 years, most of the species are confined to $\left[-1600, 1400\right]~\kmps$, with only low mass C+O tail extending beyond this range. 
Therefore, the reverse shock reduces the velocity extent in both viewing directions and makes them more symmetrical by suppressing the plumes, but its observational signature depends on orientation: in \textit{LOS-max}, it shifts and weakens a high-velocity plume component while preserving a larger positive-velocity extent for the metal-rich ejecta, whereas in \textit{LOS-min}, it mainly narrows a central peaked metal-rich ejecta distribution.  

\begin{figure*}[htbp]
  \gridline{
    \fig{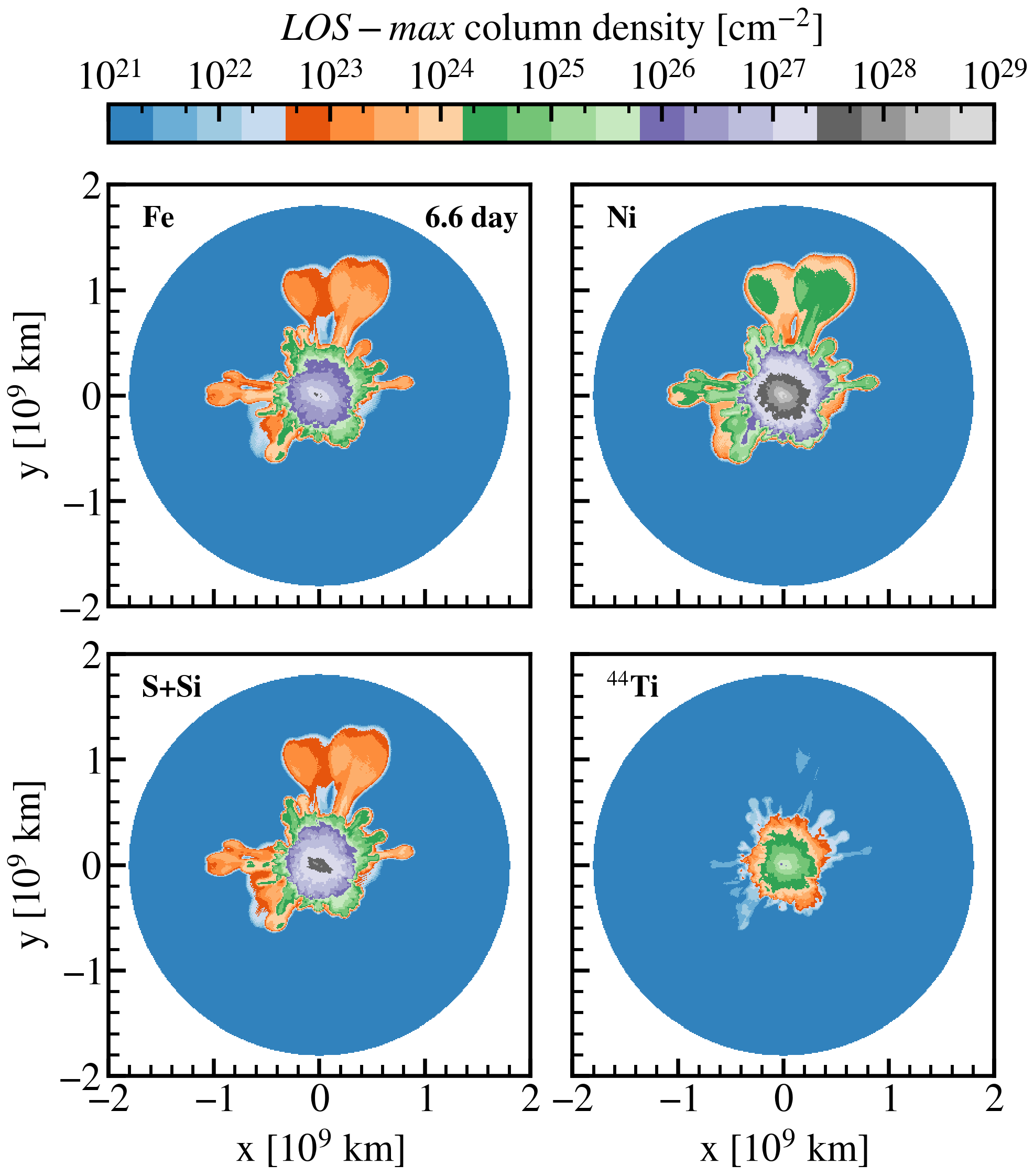}{0.48\textwidth}{(a)}
    \fig{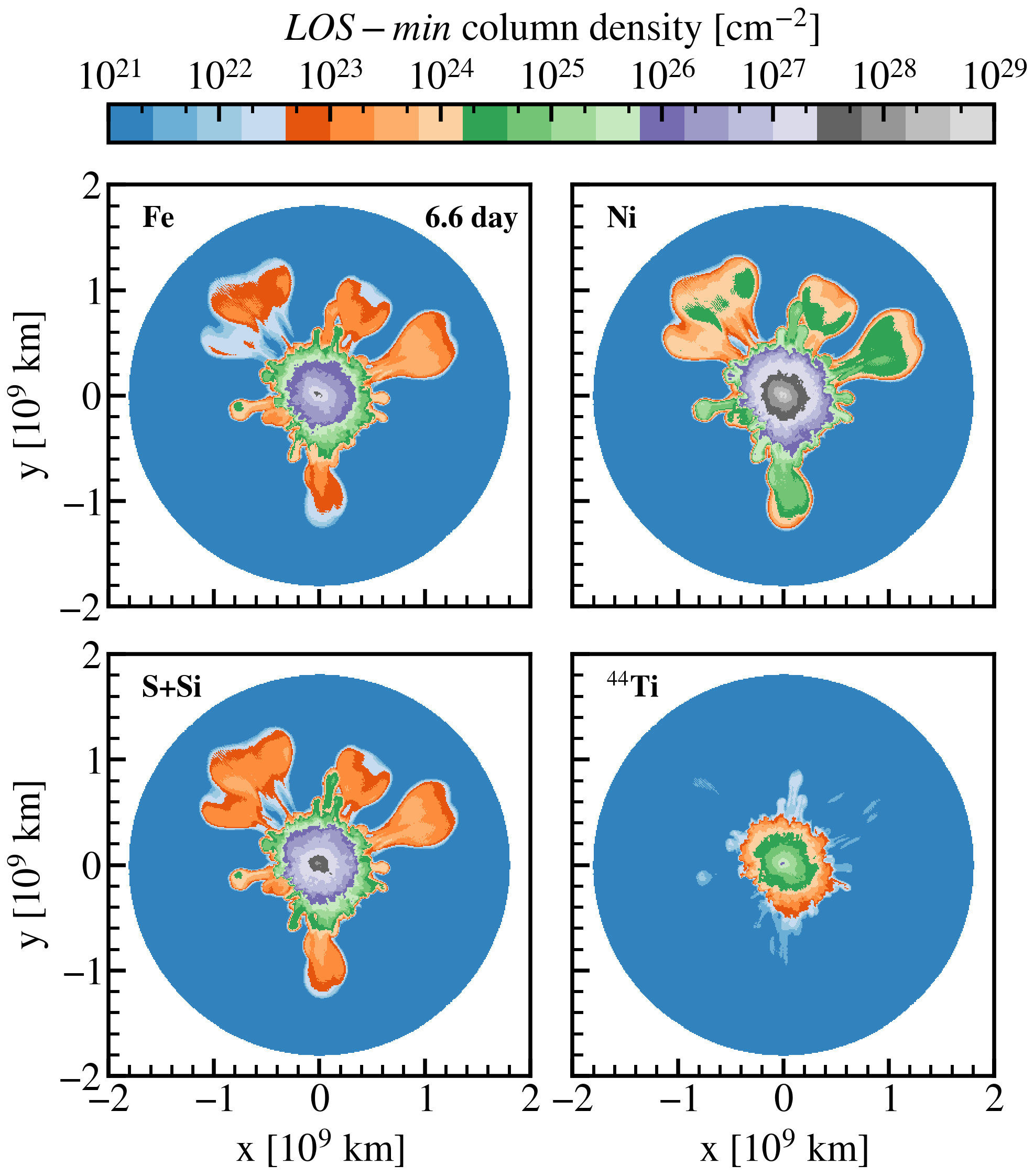}{0.48\textwidth}{(b)}
  }
  \gridline{
    \fig{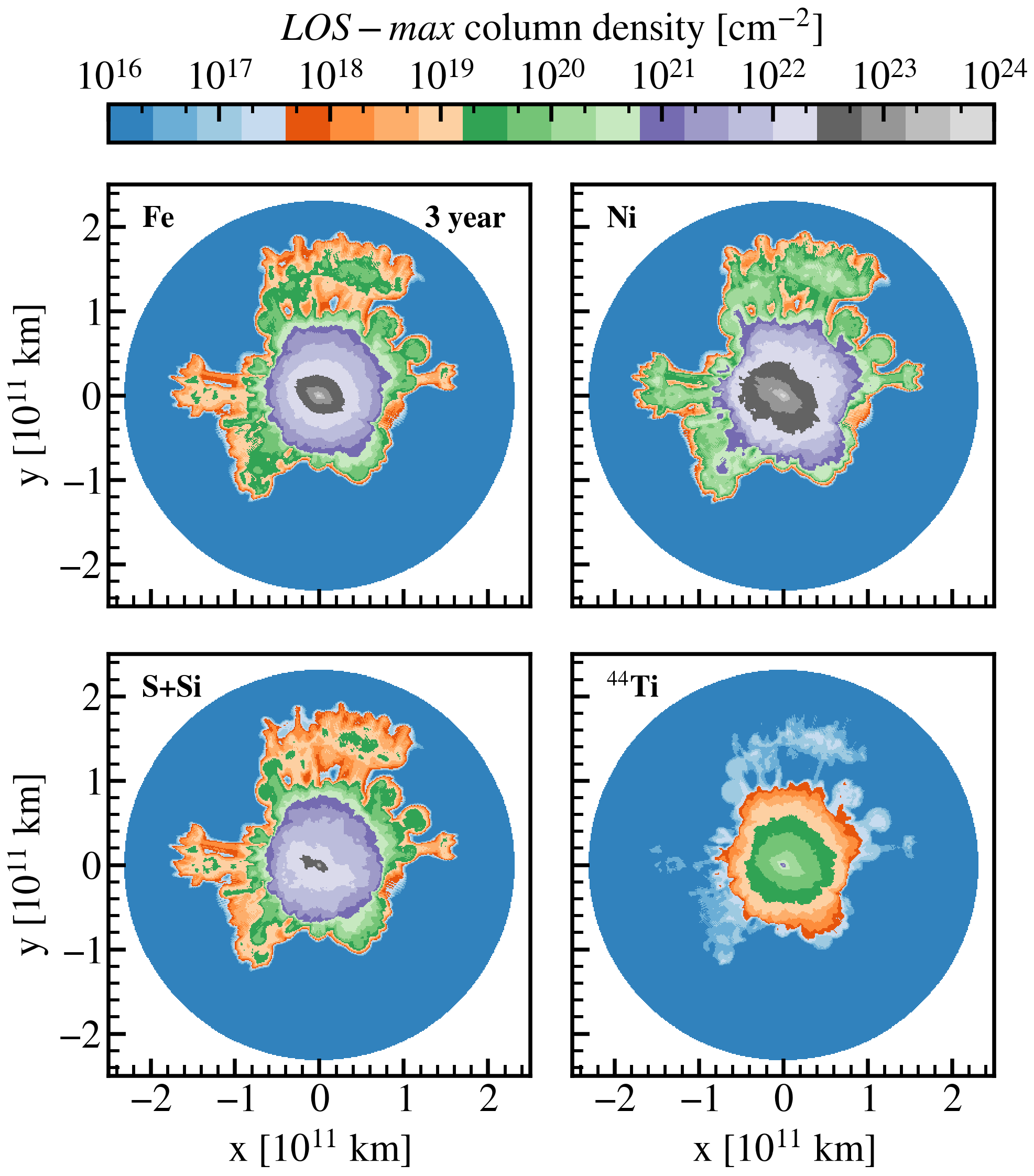}{0.48\textwidth}{(c)}
    \fig{./LOS-max_elemental_columndensity_1383_scaled.png}{0.48\textwidth}{(d)}
  }
  \caption{Column densities of Ni, Fe, S+Si, and \isotope{Ti}{44} projected along two LOS at two different times. Top left $2 \times 2$ grid (a) shows the column densities along \textit{LOS-max} at 6.6~day. Top right $2 \times 2$ grid (b) shows the column densities along \textit{LOS-min} at 6.6~day. Bottom left $2 \times 2$ grid (c) shows the column densities along \textit{LOS-max} at 3~year. Bottom right $2 \times 2$ grid (d) shows the column densities along \textit{LOS-min} at 3~year.}
  \label{fig:los-combined}
\end{figure*}

The corresponding projected column number densities of several element and \isotope{Ti}{44} are shown in Figure~\ref{fig:los-combined}. 
The dependence of projected morphology on viewing angle is clear by comparing Figures~\ref{fig:los-combined}(a)~and~\ref{fig:los-combined}(b).
In Figure~\ref{fig:los-combined}(a), which is integrated along \textit{LOS-max}, the two largest plume structures are viewed along their axes. 
As a result, much of the plume material is projected on top of the inner ejecta. 
Since the inner ejecta has column densities several orders of magnitude larger than the extended plume material, the plume morphology is masked in this viewing direction, leading to a morphology with extended features in the northern hemisphere. 

The projections for different species shown in Figure~\ref{fig:los-combined}(a) exhibit only modest geometric differences when compared to each other.
The global dominance of Ni over Fe in this model is represented in both the central region and each of the visible plumes, with nickel typically an order of magnitude more abundant than iron and the sum of silicon and sulfur.
At this point in time, 6.6 days after the nucleosynthesis completed, the \isotope{Ni}{56}, which will eventually contribute to iron as \isotope{Fe}{56}, has only decayed to \isotope{Co}{56}.
The column density of S+Si is generally comparable to iron or modestly higher. 
Visually, \isotope{Ti}{44} seems to be an exception to the compositional homogeneity in the metal-rich regions, present in the core but disappearing from the plumes.
However, the column density of \isotope{Ti}{44} is roughly 3 orders of magnitude less than nickel in both the core and the plumes. 

In contrast, Figure~\ref{fig:los-combined}(b), which is integrated along \textit{LOS-min}, provides a clearer view of the large scale plume morphology.
Since \textit{LOS-min} is approximately normal to the plane containing the three largest plumes, the dominant plume structures are well represented in the plane of the projection. 
The metal-rich ejecta appear more extended and multi-lobed in projection, especially for Ni, Fe, and S+Si. 
Features that are masked in the \textit{LOS-max} are more clearly separated from the central high-column density region in \textit{LOS-min}. 
The chemical homogeneity discussed above is also seen here.
There is however one plume, at roughly 10 o'clock in Figure~\ref{fig:los-combined}(a) which is relatively rich (by an order of magnitude) in Si+S compared to Fe, however the Ni density is larger than both.

At $t \sim 6.6$~days, the projected ejecta are still compact, extending to $r < 1.25 \times 10^{9}$~km, and the column densities span roughly $10^{21}$ to $10^{29}$~\colden\ in both \textit{LOS-max} and \textit{LOS-min} directions.
The Fe, Ni, and S+Si number density images show very similar large-scale structures, indicating that these species are dynamically coupled and are carried outward by the same plume morphology. 
Their largest column densities are concentrated near the inner ejecta, reaching $\sim 10^{26}$ to $10^{29}$~\colden, while the extended plume material has lower column densities of $\sim 10^{22}$ to $10^{25}$~\colden.
The contrast between the high-column inner ejecta and the lower-column plume material is what makes the viewing angle dependence strong.

The same viewing angle dependence persists until the end of the simulation at 3.0~years, as shown in Figures \ref{fig:los-combined}(c)~and~\ref{fig:los-combined}(s). 
By 3.0~years, the supernova ejecta has expanded to radii $r < 2.0 \times 10^{11}$~km, causing the column densities to decrease by several orders of magnitude, reaching only $\sim 10^{21}$ to $10^{24}$~\colden\ for the inner elemental ejecta and $\sim 10^{16}$ to $10^{20}$~\colden\ for the outer plumes. 
By 3.0 years, the reverse shock has interacted with the extended plumes, considerably decelerating them.
As a result, the inner ejecta appears relatively larger in Figures \ref{fig:los-combined}(c)~and~\ref{fig:los-combined}(d).
Interaction with the reverse shock has also added considerable smaller scale structure to the plumes.
Nevertheless, the interaction with the reverse shock does not make the ejecta smooth or spherical in either projected plane and they continue to show the large-scale asymmetries and retain the imprint of the plume geometry established at shock breakout.  

Observationally, these maps demonstrate the need to interpret the projected morphology and the LOS velocity structure together. 
When the structure has a few dominant plumes, a viewing angle direction aligned with these plumes can produce a broad LOS velocity distribution while making the projected ejecta appear more compact. 
Conversely, a viewing angle perpendicular to the plume plane can produce a narrow LOS velocity distribution while revealing a more extended and multi-lobed morphology. 
Therefore, the inferred degree of mixing, the observed asymmetry, and even the visibility of individual plume features can strongly depend on viewing angle.

\subsection{Global Evolution}

\begin{figure}
    \centering
    \includegraphics[width=\linewidth]{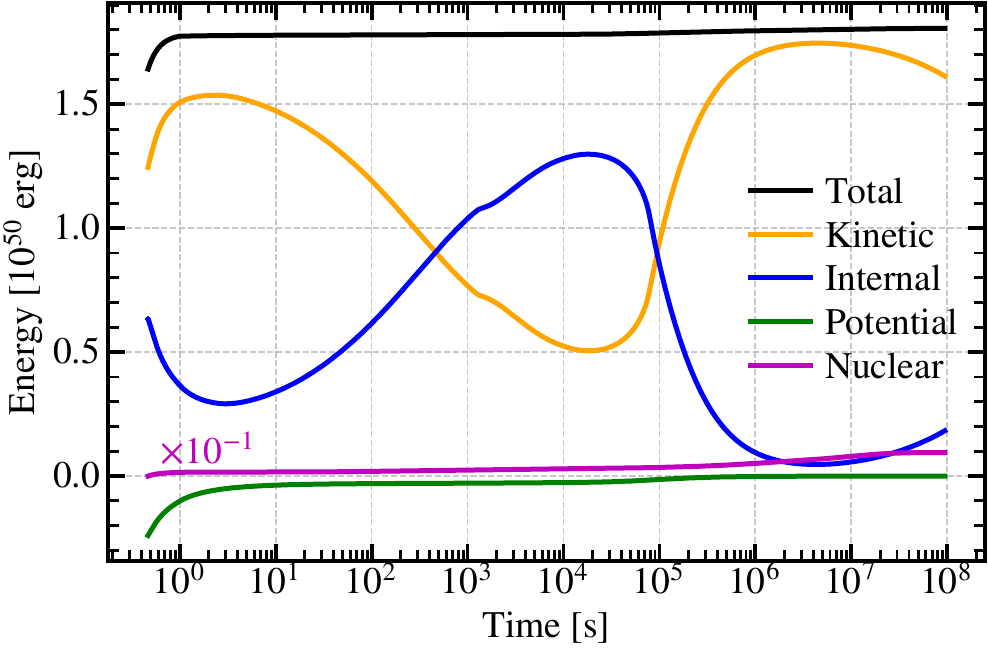}
    \caption{Evolution of energy in the 3D D9.6 model. Nuclear energy curve shows cumulative decay energy scaled by 10 for visibility.}
    \label{fig:global_energetics}
\end{figure}

The acceleration and deceleration of the shock as it moves through the progenitor star and into the CSM, shown in Figure~\ref{fig:avg_shock_velx}, also affects the global energy evolution shown in Figure~\ref{fig:global_energetics}. 
Note the slight increase in total energy and potential energy beyond $10^{4}$~s, which primarily results from the continued expansion of the ejecta. While after $10^{6}$~s, the acceleration of the CSM material due to the shock primarily continues to increase the total energy.

At the time of mapping to Flash-X, the shock is in a region with a negative $\rho r^{3}$ gradient, and therefore the shock accelerates. 
This results in an increase in kinetic energy, while the expansion causes the internal energy to decrease. 
Further, the NS wind adds energy to the system, which results in the steady increase of total energy until 1.0~s, at which point the NS wind subsides.
As the shock propagates through the C+O layer into the He layer, it enters a region with a positive $\rho r^{3}$ gradient, beginning at $r \sim 1.0 \times 10^{4}$~km. 
In this region, the shock decelerates and compresses the post-shock material, increasing the internal energy at the expense of kinetic energy. 
The positive $\rho r^{3}$ gradient persists at the He/H interface, and hence the internal energy continues to increase.

At the He/H interface, a weak second reverse shock forms, which steepens over time and further compresses the inner ejecta. 
This compression contributes to the increase in internal energy, which reaches a peak value of about $1.30\times10^{50}$~erg at about $t \approx 19000$~s, as shown in Figure~\ref{fig:global_energetics}. 
This conversion continues until the shock reaches the edge of the progenitor at $r \sim 1.5 \times 10^{8}$~km, where it passes through a steep negative gradient in $\rho r^{3}$, causing the shock to rapidly accelerate. 
This expansion converts internal energy back into kinetic energy. 
The momentum carried by the shock, together with energy deposited by radioactive decay heating from $\isotope{Ni}{56} \rightarrow \isotope{Co}{56} \rightarrow \isotope{Fe}{56}$ chain (hereafter \isotope{Ni}{56} chain), contributes to the continued increase in kinetic energy until 50~day, when it reaches a maximum value of about $1.75 \times 10^{50}$~erg, almost equal to the total energy.

By this time, the shock has swept up more CSM mass than the ejecta mass, thus the CSM interaction becomes dynamically important. 
The forward shock begins to decelerate, while the reverse shock formed near the H/CSM interface compresses and decelerates the leading ejecta plumes.
This interaction converts part of the ejecta kinetic energy into internal energy through shock heating and compression, leading to the decrease in kinetic energy and the corresponding increase in internal energy seen at late times in Figure~\ref{fig:global_energetics}. 

Describing the energy of a supernova is traditionally a point where observations and theory do not speak the same language.
Observers estimates are based on the kinetic energy of the ejecta, while modelers, stopping their models after just a few seconds, construct a diagnostic energy by summing the kinetic, thermal and gravitational energies over their ejecta.
As Figure~\ref{fig:global_energetics} demonstrates, these measures are only equivalent after shock breakout and only for a period of weeks before interaction with the CSM again converts kinetic energy into thermal energy.

\begin{figure*}
    \centering
    \includegraphics[width=\linewidth]{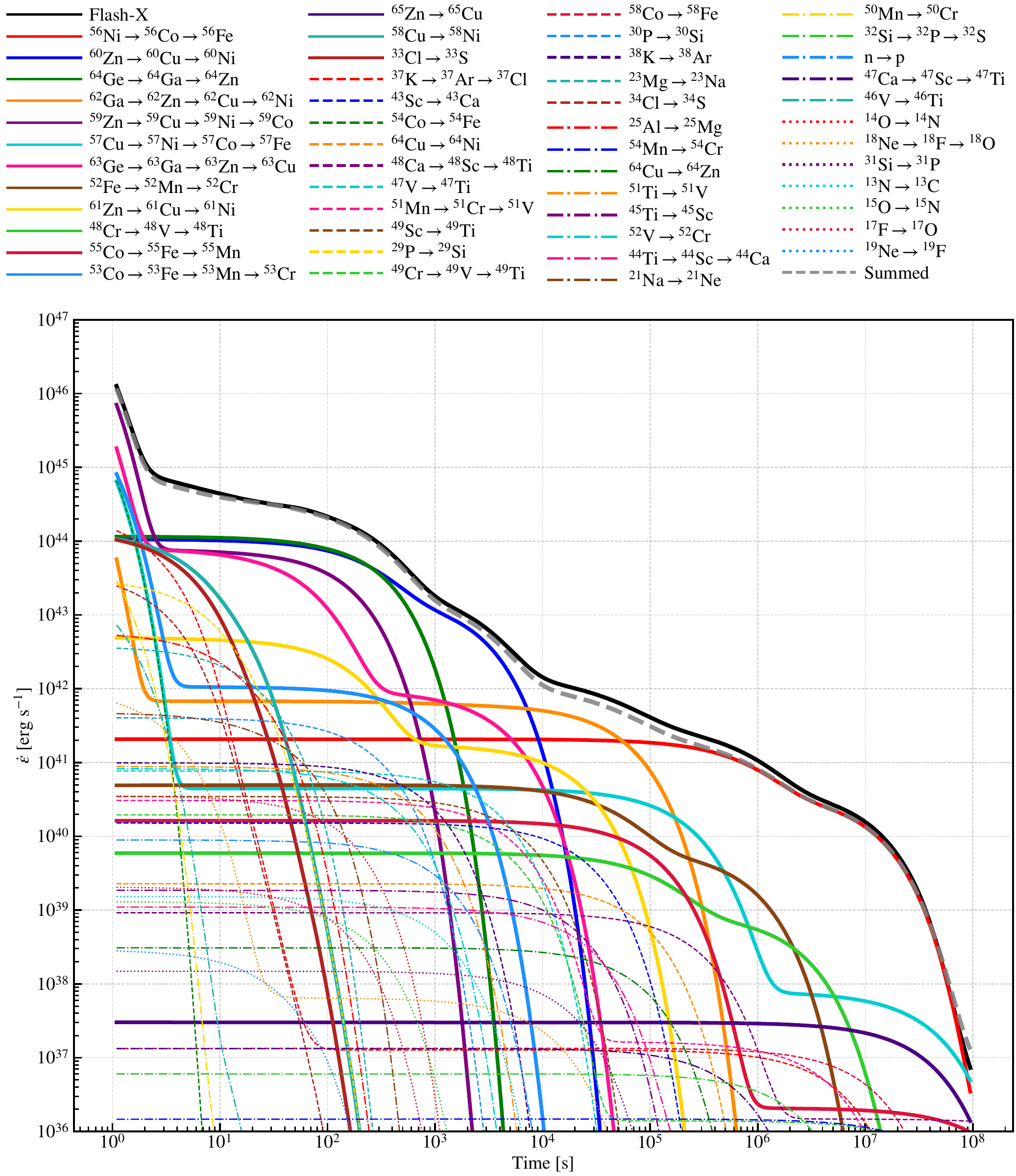}
    \caption{Decay energy generation rate of radioactive isotopes during 3D D9.6 model evolution. Black line is energy generation rate obtained directly from the simulation. Thick gray dashed line is summed rate from all decay chains. Decay chains with maximum rates less than $10^{36}$~$\mathrm{erg~s}^{-1}$ have been omitted for clarity.}
    \label{fig:decay_energy_plot}
\end{figure*}

The inclusion of a 160-isotope nuclear reaction network in our simulations allows us to accurately capture the nuclear energy generation during the explosive burning phase, while also providing a much more complete description of the radioactive decay energy budget at later times. 
In particular, this enables us to follow not only the canonical $\isotope{Ni}{56}$ decay chain, but also the many subdominant decay chains that contribute at different stages of the evolution. 
Figure~\ref{fig:decay_energy_plot} shows the radioactive heating rate as a function of time for the decay chains present in our network. 
The black curve shows the total radioactive heating rate computed directly by Flash-X during the simulation, while the individual decay chain contributions are obtained by post-processing simulation checkpoint files.
The ordering of lines in Figure~\ref{fig:decay_energy_plot} reflects the total energy contribution of that decay chain in the first 3 years of the simulation.
Because the checkpoint files are written much less frequently than the hydrodynamic time steps and become increasingly widely spaced as the simulation progresses, the decomposition does not fully resolve rapid changes in the decay power of short-lived isotopes as well as the mass that is lost when blocks are dropped.  
Even with this limitation, the sum of the individually tracked decay chains (represented by a gray thick dashed line) closely reproduces the total heating rate released by the full decay network. 
At later times, however, the surrounding CSM out to the edge of the computational domain becomes increasingly transparent to decay $\gamma$-rays. This becomes more prominent for  $t \ge 4 \times 10^7$~s, as shown in Figure \ref{fig:decay_energy_plot}, where the summed energy generation rate reconstructed from the individual decay chains exceeds the nuclear energy generation rate deposited in the simulation. This difference reflects the increasing escape of decay $\gamma$-rays. By the end of the simulation, approximately 65\% of the energy generated by radioactive decay is escaping the domain.  
The early decay power is distributed among several short-lived species, while at later times the contribution becomes increasingly dominated by the longer lived chains.  
Despite the relatively small production of \isotope{Ni}{56} in this ECSN-like supernova, the \isotope{Ni}{56} decay chain remains the single most important contributor, accounting for 75.6\% of the total decay energy released by the radioactive isotopes over the simulation. 
The remaining 24.4\% comes from other decay chains, particularly \isotope{Zn}{59}, \isotope{Ge}{64}, \isotope{Zn}{60},  and \isotope{Ga}{62}, in the first hours of the explosion and \isotope{Cu}{57} and \isotope{Zn}{65} in later years (see Figure \ref{fig:decay_energy_plot}).
This highlights the impact of evolving a large nuclear network when modeling the long-time radioactive heating of the ejecta, especially for a low-mass progenitor.
This relatively large non-\isotope{Ni}{56} chain contribution is a direct result of the  electron-capture supernova (ECSN)-like nature of this supernova, providing considerable neutron-rich ejecta that would not be found in most CCSN.

\section{Summary and Conclusions}

We have presented a series of long-time CCSNe simulations of a 9.6-\msun\ zero-metallicity progenitor using the Flash-X code. 
These simulations begin near the end of the neutrino-driven explosion phase, building on a neutrino radiation hydrodynamic  simulation, and follow the evolution of the shock and ejecta through the progenitor star, shock breakout, and expansion into a steady wind CSM. 
This work builds directly on the simulations of \citetalias{Sandoval_2021}, which evolved the same progenitor with a 160-isotope nuclear reaction network but stopped just before shock breakout.
Here, we extend the evolution to multi-year timescales and include parametrized NS cooling and wind, a steady-wind CSM, and a large radioactive decay network. 
These suite of 1D, 2D, and 3D calculations also allows us to isolate how each added physical ingredient modifies the ejecta evolution, and how those effects change once the evolution becomes multidimensional. 
In particular, the detailed composition inherited from the \chimera\ explosion model allows us to follow electron capture and $\beta$-decay chains with a 160-isotope decay only network, rather than limiting the radioactive heating to the canonical $\isotope{Ni}{56}$ decay chain. 
This framework allows us to connect the early explosion asymmetries inherited from the neutrino-driven phase to the multi-year evolution of the ejecta morphology, velocity structure, projected column densities, and radioactive decay energy budget.

In 1D, we performed a sequence of six simulations designed to isolate the effects of enhanced physics. The baseline model, \runlabel{1}{DR}{1}, extends a setup similar to \citetalias{Sandoval_2021} beyond shock breakout by including a CSM. 
As the shock propagates through the He/H interface and then into the CSM, it produces a reverse shock, a forward shock, and a contact discontinuity where shocked ejecta and shocked circumstellar material accumulate into a narrow density shell. 
The NS cooling and accretion only have a modest effect on the global shock evolution, mainly reducing the density of the innermost ejecta. 
The NS wind has the largest effect on the shock position among the added early-time physics, pushing the ejecta and shock slightly farther outward. 
Radioactive decay heating produces the most significant change in the inner density structure: it forms a hot bubble, lowers the interior density by more than an order of magnitude, and suppresses the inward progress of the reverse shock. 
The assumed 20\% neutrino loss weakens this effect slightly, while $\gamma$-ray escape has little dynamical effect in the 1D runs because the ejecta remains optically thick during the evolution.

In 2D, we compared two simulations, the baseline model \runlabel{2}{DR}{1} and the full physics model \runlabel{2}{DR}{6}. 
Due to the higher explosion energy of the 2D \chimera\ model when compared to 1D, the 2D results cannot be solely attributed to the dimensional extension of the 1D sequence. 
However, they show the effect of the same added physics when the radioactive species are spatially localized. 
Rather than producing the nearly uniform low-density central bubble in 1D, decay heating is spatially localized to the radioactive metal-rich ejecta pockets. 
These pockets expand, lower their densities, and compress the surrounding material, leading to the formation of dense shells separating the decay heated and non-decay heated regions. 
In this case, the high-density shell behind the shock become the base of RT plume growth, and the metal-rich bubble exceeds farther outward in \runlabel{2}{DR}{6}, than in \runlabel{2}{DR}{1}. 
However, similar to the 1D case, the ejecta for both of the models cannot catch up to the shock front and lags behind the main shock front significantly.

In 3D, we evolved the full physics model from 466.6~ms to 3~years post-bounce. 
The overall evolution prior to the shock breakout is broadly consistent with \citetalias{Sandoval_2021}, but the additional physics changes the detailed plume morphology. 
The NS wind initially stalls accretions and pushes the inner ejecta outward, while the later radioactive heating inflates the metal-rich structures. 
The push of the inner ejecta due to the NS wind keeps the high density material closer to the shock front and results in the formation of additional RT plumes. 
Compared with the earlier FLASH calculations of the same progenitor, our model produces fewer small clumps and more large structure. 
This is consistent with the decay heating inflating and merging smaller metal-rich plumes, although the differences cannot be solely attributed to the decay heating as the modes also differ in NS physics and the numerical hydrodynamics scheme. 
The presence of asymmetric elongated plumes, some of which also penetrate the main shock front, results in a asymmetrical shock breakout, which starts at around 68000~s and continues until 80000~s, spanning more than 3~hours. 

After shock breakout, the shock accelerates through the steep density gradient at the progenitor surface and enters the CSM, resulting in the formation of a third reverse shock. 
The leading plumes accelerate behind the shock front, but cannot keep up with them. These plumes reach the reverse shock first and start to get deformed by 20~days. 
While the plumes continue to get progressively shredded by the reverse shock, they continue to exhibit their large-scale asymmetry. 
The radial velocity evolution of \isotope{Z}{56} mass fractions, shows the effects of decay heating and the interaction with the reverse shock. 
The high-velocity plume component peaks at 2200~\kmps\ bin at approximately 6 days, but is decelerated by reverse shock to about 1400~\kmps\ bin by 3~years. 
In contrast, the inner metal-rich material with a peak velocity of 350~\kmps\ bin by 1~day continues to broaden and reaches a peak velocity of about 650~\kmps\ bin by the end of the simulation, likely due to radioactive heating. 
The 160-isotope network also shows that the canonical $\isotope{Ni}{56}$ chain dominates the radioactive energy budget, but does not account for all of it, with 24.4\% of the radioactive energy coming from other decay chains. 

There is also a strong viewing angle dependence of the observational diagnostics. 
When the ejecta is viewed along the direction of propagation of the strongest plumes, it show a broader line of sight velocity distribution but a more compact projected morphology because the plume material is projected against the high column density inner ejecta. 
However, when the dominant plume motion lies transverse to the line of sight, the observed line of sight velocity distribution is narrower, but the projected column density maps reveal a more extended and multi-lobed plume structure. 
This shows that projected morphology and velocity structure must be interpreted together: a compact image does not necessarily imply weak mixing, and a narrow line profile does not necessarily imply spherical ejecta. 
In both viewing directions, Fe, Ni, and S+Si largely trace the same plume structures. The \isotope{Ti}{44} distribution also shows weak extensions along these plumes, however, its column density is about three to four orders of magnitude lower than Ni in the plumes. As a result, \isotope{Ti}{44} appears more compact and centrally concentrated in the projection. 
By 3~years, expansion lowers the column densities by several orders of magnitude and the reverse shock narrows the velocity distributions, but the ejecta remain globally asymmetric and retains the imprints of the early plume geometry.

Our results are qualitatively consistent with the \cite{gabler_2020} long-time ejecta simulations of more massive (15- and 20-\msun) progenitors. 
Similar to their case, we observe that the decay heating inflates the metal-rich ejecta, produces low-density radioactive bubbles, and compresses the surrounding material into denser shells. 
However, there are important difference between the two calculations. 
While \cite{gabler_2020} focus on the canonical $\isotope{Ni}{56}$ decay chain, along with the $\isotope{Ti}{44}$ decay chain, the 160-isotope decay network in our simulation accounts for many more electron capture and $\beta$-decays. 
These additional chains contribute most strongly at earlier times and can help modify the plume morphology by inflating and merging metal-rich structures. 
In our simulation, radioactive inflation also acts together with reverse shock interaction after breakout, so the late-time morphology reflects both decay heating of the metal-rich ejecta and hydrodynamic deceleration by the reverse shock in the CSM. 

This comparison with \cite{gabler_2020} also highlights the progenitor dependence of the long-time ejecta evolution. 
Their more massive progenitors result in CCSN with stronger compositional asymmetries than our very low mass progenitor. 
They also produce substantially more \isotope{Ni}{56}, leading to a stronger \isotope{Ni}{56}-bubble effect and different late-time ejecta structure.
This sensitivity to progenitor and explosion properties is also evident in the long-time simulations that connect the explosion to the remnant phase.  
Simulations built to match SN~1987A and Cas~A \citep{Orlando2020, Orlando2021} show that the final remnant morphology depends on the progenitor structure, explosion asymmetry, radioactive plume evolution, reverse-shock interaction, and the surrounding medium.

A similar dependence is seen even among simulations of the same 9.6~\msun\ progenitor, highlighting the qualitative but not quantitative agreement between CCSN codes.
\citet{stockinger2020} evolved this progenitor through shock breakout and fallback with a lower explosion energy than in the present calculation, finding relatively spherical shock evolution and limited mixing of \isotope{Ni}{56} into the bottom of the H envelope.
The 3D simulation of the same progenitor by \citet{WangBurrows_2024} has a higher explosion energy and larger \isotope{Ni}{56} yield than our model, while also finding substantial neutron-rich ejecta.
Thus, the detailed shock evolution, plume morphology, ejecta velocities, radioactive heating budget, and late-time remnant structure are tied to the explosion properties and to the density and composition structure of the progenitor.
A larger set of long-time 3D simulations will be required to determine how these quantities vary across low-mass progenitors and how they connect to the broader population of CCSNe.

The energetics and nucleosynthesis yields of our simulation are consistent with several observational diagnostics proposed for ECSNe by \citet{Hiramatsu_ecsn_2021}. 
The low explosion energy of $1.68 \times 10^{50}$~erg, the low \isotope{Ni}{56} yield of $\sim 2.6 \times 10^{-3}$~\msun\ during explosive burning, and the enhanced $\left[\mathrm{Ni}/\mathrm{Fe}\right] \approx 1.7$ at 3~years, suggestive of neutron-rich ejecta, are all consistent with an ECSN-like event. 
Therefore, although our progenitor is a low-mass Fe-core CCSN model rather than an ONeMg-core ECSN progenitor, its energetic and nucleosynthesis signatures place it in the same broad observational regime as the ECSN-like interpretation of SN 2018zd \citep{Itagaki_SN2018zd}.

Beyond these global explosion diagnostics, the morphology and composition of the ejecta provide an additional way to compare our model with observed remnants. 
In particular, our model does not appear to be a close analog of Cas~A, since there are not strong inhomogeneities in the metals composition \citep{Grefenstette_2017,Grefenstette_2014,HuRaBu00}. 
Similarly, the strong compositional asymmetries seen in G292.0+1.8 \citep{BhPaSc19,PaHuSl07} are inconsistent with this ECSN-like supernova explosion.
Instead the metal-rich ejecta, while localized in the H-rich remnant, has fairly uniform consistency, likely as a result of the much more spherical explosion and the relatively small role that neutrino-driven convective plumes have in the launching for the explosion in the seconds after the formation of the neutron star.
Recent JWST observations of SNR~0540-69.3 show inner ejecta are dominated by two highly fragmented lobes, with different emission lines tracing different 3D morphologies \citep{Larsson_2026}. This is in contrast to our model which has nearly spherical inner ejecta, with different ejecta tracing similar morphology, making the SNR unlikely to have originated from a low mass ECSN-like core collapse supernovae.

\begin{acknowledgments}
We thank Sabrina DeSoto for helpful comments during the preparation of this manuscript.
S.N. was supported by the National Science Foundation Nuclear Theory Program (PHY-2309988, PHY-1913531). 
W.R.H. acknowledges support from the U.S. Department of Energy, Office of Nuclear Physics. 
O.E.B.M. acknowledges support from the U.S. Department of Energy, Office of Advanced Scientific Computing Research. 
This research used resources of the Oak Ridge Leadership Computing Facility at Oak Ridge National Laboratory. 
Research at Oak Ridge National Laboratory is supported under Contract No. DE-AC05-00OR22725 from the Office of Science of the U.S. Department of Energy to UT-Battelle, LLC.

\end{acknowledgments}

%
\facilities{OLCF}

\software{Flash-X \citep{dubey2022}, VisIt \citep{Childs_High_Performance_Visualization--Enabling_2012}, Matplotlib \citep{Hunter2007}, NumPy \citep{numpy_cite}, h5py \citep{h5py_cite}, SciPy \citep{scipy_cite}}


\bibliography{apj-jour,add_journals,citation}{}
\bibliographystyle{aasjournalv7}



\end{document}